\documentclass{article}

\usepackage{arxiv}

\usepackage[utf8]{inputenc} 
\usepackage[T1]{fontenc}    
\usepackage{hyperref}       
\usepackage{url}            
\usepackage{booktabs}       
\usepackage{amsfonts}       
\usepackage{nicefrac}       
\usepackage{microtype}      
\usepackage{lipsum}		
\usepackage{graphicx}
\usepackage[numbers]{natbib}
\usepackage{doi}
\usepackage{longtable}
\usepackage{colortbl}
\usepackage{caption}
\usepackage{multirow}

\title{An Exploration of Geo-temporal Characteristics of Users' Reactions on Social Media During the Pandemic}

\author{Eldor Abdukhamidov \\
	Department of Computer Science and Engineering\\
	Sungkyunkwan University\\
	South Korea \\
	\texttt{abdukhamidov@skku.edu} \\
	\And
	Firuz Juraev \\
	Department of Computer Science and Engineering\\
	Sungkyunkwan University\\
	South Korea \\
	\texttt{fjuraev@g.skku.edu} \\
	
	\And
	Mohammed Abuhamad \\
	Department of Computer Science \\
	Loyola University Chicago\\
	Chicago, USA \\
	\texttt{mabuhamad@luc.edu} \\
	
	\And
	Tamer AbuHmed \\
	Department of Computer Science and Engineering\\
	Sungkyunkwan University\\
	South Korea \\
	\texttt{tamer@skku.edu} \\
	
}

\renewcommand{\headeright}{}

\begin{document}
\maketitle

\begin{abstract}
During the outbreak of the COVID-19 pandemic, social networks become the preeminent medium for communication, social discussion, and entertainment. Social network users are regularly expressing their opinions about the impacts of the coronavirus pandemic. Therefore, social networks serve as a reliable source for studying the topics, emotions, and attitudes of users that are discussed during the pandemic. In this paper, we investigate the reactions and attitudes of people towards topics raised on social media platforms.
We collected data of two large-scale COVID-19 datasets from Twitter and Instagram for six and three months, respectively. The paper analyzes the reaction of social network users on different aspects including sentiment analysis, topics detection, emotions, and geo-temporal characteristics of our dataset. We show that the dominant sentiment reactions on social media are neutral while the most discussed topics by social network users are about health issues. The paper examines the countries that attracted more posts and reactions from people, as well as the distribution of health-related topics discussed in the most mentioned countries. We shed light on the temporal shift of topics over countries. Our results show that posts from the top-mentioned countries influence and attract more reaction worldwide than posts from other parts of the world.
\end{abstract}


\section{Introduction}
The COVID-19 has been spread rapidly around the world since December 2019. The disease is considered a pandemic that impacted many countries across all living continents. Due to its consequences (infection, increasing death rate), governments developed policies to decrease the virus's spread. Quarantine, social distancing are among the measures taken by governments as emerging actions. Such measures have led to stopping social events, starting online lectures for pupils, college, and university students, reducing working hours at workplaces, and turning into telecommuting. As a result, social networks have become the main platform for people to express opinions and share information. Social media, such as Instagram and Twitter, has grown due to measures like social distancing.

Recently, various research works have been conducted on COVID-19 \cite{Shuja2020} related issues, including spread modeling \cite{Shrivastav2020}, cases prediction \cite{Kumari2020,Singh_2020}, social networks reactions {\cite{Lamsal2020}}, diagnosis \cite{Goel2020,Turkoglu2020,Li_2020,Mukherjee2020}, etc. This paper presents analyses of how users react on Twitter and Instagram on posts and topics discussed during the pandemic. Our analyses include sentiment study on topics and posts from the two social media platforms, in depth-analysis of geographical distribution and influence traits, and investigating geo-temporal characteristics of people's reactions. To this end, we collect and analyze two large-scale datasets from Twitter and Instagram. We utilize state-of-the-art Natural Language Processing (NLP) methods to derive linguistic annotations for the sentiment of COVID-19 related tweets and posts. This step includes topic modeling using Latent Dirichlet Allocation (LDA) on the datasets with different sentiment types (i.e., very negative, negative, neutral, positive, very positive). The adopted topic modeling approach enables discovering topics discussed by people on the social platforms and the per-topic sentiment analysis. We also employed Named Entity Recognition (NER) methods to identify named entities from tweets and posts related to geographical locations, such as countries. For our analysis, the top-5 positively mentioned countries are selected for further investigation regarding discovering the geo-temporal characteristics of people’s reactions. Moreover, we explore the spikes of sentiments on posts from the top-mentioned countries and study the topics during these spike periods. 
In addition, we analyze the geo-temporal characteristics using state-of-the-art NLP-based methods to observe the temporal shift in the frequency of posts and topics and the reactions towards posts from the top-mentioned countries through time. 

Our study aims to answer the following main research questions: (1) What are the major topics that people discuss during the pandemic and their reactions to these topics? (2) What are the geographical characteristics of posts on social networks, and whether there are localities that are more influenced by the pandemic than others? (3) Focusing on the top-mentioned and most-influential geographical localities, what are the major discussed topics? (4)  What are the geo-temporal traits observed on topics during the pandemic regarding whether they shift, diminish, or persist during the observation period? 
To sum up, the summary of our contributions are as follows.

\begin{itemize}
\item We collected a large-scale dataset of users’ posts and tweets from Twitter and Instagram.
\item Sentiment analysis has been conducted for social networks based on five sentiment categories: very negative, negative, neutral, positive, and very positive.
\item Topic modeling analysis on the sentiment categories was carried out to identify overall public reactions to COVID-19 related tweets and posts.
\item Exploration of geo-temporal characteristics based on countries was performed to recognize countries that are mentioned in COVID-19 tweets and posts (refer them as top countries).
\item We also performed sentiment analysis of top countries to determine people’s emotional responses about mostly mentioned countries.
\item Exploiting the applied topic modeling technique, we extracted the major discussed topics on the top mentioned countries in our dataset.
\item An analysis based on word2vec embedding was conducted to discover the shift of words and topics used with top countries over time.
\item The study provided locality analysis of top countries to find countries where tweets and Instagram posts about top countries are published.
\end{itemize}

The rest of the paper is organized as follows. Section 2 explores the related work on studying people's reactions on social networks during the COVID-19 pandemic. Section 3 presents the methods used for the data collection and a description of the collected data. Section 4 shows our sentiment analysis on the collected dataset. The per-sentiment topic modeling is described in Section 5. We provide the geo-temporal analysis on the collected dataset in Section 6 and conclude in Section 7.

\section{Related works}

Since January 2020, several studies have explored the impact of COVID-19 on people’s daily lives. In this section, we review the main related work in social networks analysis for COVID-19. The work of Schild et al. \cite{Schild2020} was among the first studies on analyzing people’s Sinophobic behavior on social media. Their work included studying two datasets of posts from Twitter and 4-chan’s (/pol/) that were published between November 1, 2019, and March 22, 2020. Both datasets were used to explore whether there was an important change regarding the spreading of Sinophobic contents among social network users. The authors trained three word2vec models over the social network data to study the use of words and the major differences in using the most similar words, and to observe the development of new terms. The authors observed that COVID-19 caused the emergence of new Sinophobic slurs and a trend of blaming China for the COVID-19 pandemic. 

Ordun et al. \cite{Ordun2020} proposed a paper to answer prevalent COVID-19 related questions about trends, news-making events, topics, retweets, and COVID-19 networks, by analyzing 23,830,322 tweets posed during the time period between March 24, 2020, and April 9, 2020. The authors revealed the top trends in tweets using Keyword Trend Analysis. They adopted a topic modeling stage using Latent Dirichlet Allocation (LDA) to spot the events that cause sparks in COVID-19 tweets. Moreover, they used Uniform Manifold Approximation and Projection (UMAP) to find unique topics. The authors also applied network modeling to derive how social media reacts to the spread of COVID-19.  Their findings pointed out the huge attention/reaction given to the live White House Coronavirus Briefings and topics related to healthcare and government reactions. We note that the authors used a pre-filtered dataset with 13 healthcare-related terms from the Twitter Streaming API and reduced the dataset size by nearly 77\% from 23,830,322 tweets to 5,506,223 tweets when removed the retweets. 

Li et al. \cite{Li2020} conducted a comprehensive analysis of datasets collected from Twitter and Weibo posted between January 20, 2020, and May 11, 2020. Six emotions containing anger, disgust, fear, happiness, sadness, surprise are identified based on the user-generated content. Authors compare people’s emotions in the United States and China to understand the different reactions towards COVID-19. Using NLP, the authors found the causes of public emotions, e.g., the reason for angriness, surprise, and worry. The study indicated a strong contrast in people’s opinions about COVID-19 in different countries. However, the study was limited to two countries, i.e., the USA and China. The authors concluded with a suggestion to use real-time emotion analysis for methods and procedures for the fight against the global crisis.  

Sharma et al. \cite{Sharma2020} proposed a dashboard for misinformation tracking on Twitter. Their work follows Coronavirus-related discussions over time based on Twitter data between March 1, 2020, and May 3, 2020, to determine the incorrect and deceptive contents. The authors conducted an analysis of public sentiments on specific data that are filtered by keywords such as “\textit{\#workfromhome}” and “\textit{\#socialdistance}”. They also analyzed topics in Twitter conversations, extracted hashtags that are emerging in different countries, and utilized tweets about countries to estimate the public perception.  However, the dataset includes tweets from only two months, limiting the insights for long-term users’ behavior on social media. 

Lamsal \cite{Lamsal2020} collected a large-scale Twitter dataset of English tweets related to COVID-19 for the period between March 20, 2020, and July 17, 2020. In their work, authors used the dataset and its filtered version that includes only geo-tagged tweets for sentiment analysis and network analysis. Tweets collected between April 24, 2020, and July 17, 2020, were utilized to study significant drops in the average sentiment over the period by generating the sentiment trend graph. The dataset that consists of only geo-tagged tweets was used to conduct network analysis; furthermore, it was used to generate the sentiment-based world and regional maps. The study identified 12 different communities based on the usage of similar hashtags within the dataset. The author also presented a set of popular hashtags and their associated communities.

\begin{table}[h]
\centering
\arrayrulecolor{black}
\captionof{table}{Comparison of related papers based on Twitter dataset and different analysis}
\label{tab:papersComparison}
\resizebox{\linewidth}{!}{%
\begin{tabular}{l|cccccccccc}
\multicolumn{1}{c|}{\multirow{2}{*}{ \textit{\textbf{Authors}} }} & \multicolumn{2}{c}{ \textit{\textbf{~Twitter}} } & \multirow{2}{*}{ \textbf{\textit{Keywords} }} & \multirow{2}{*}{\begin{tabular}[c]{@{}c@{}} \textbf{\textit{Feature }}\\\textbf{\textit{Analysis} }\end{tabular}} & \multirow{2}{*}{ \textbf{\textit{Geospatial} }} & \multirow{2}{*}{\begin{tabular}[c]{@{}c@{}} \textbf{\textit{Topic }}\\\textbf{\textit{Modeling} }\end{tabular}} & \multirow{2}{*}{ \textbf{\textit{Sentiment} }} & \multirow{2}{*}{\begin{tabular}[c]{@{}c@{}} \textbf{\textit{Locality }}\\\textbf{\textit{Analysis} }\end{tabular}} & \multirow{2}{*}{\begin{tabular}[c]{@{}c@{}} \textbf{\textit{Country }}\\\textbf{\textit{Analysis} }\end{tabular}} & \multirow{2}{*}{\begin{tabular}[c]{@{}c@{}} \textbf{\textit{Network }}\\\textbf{\textit{Models} }\end{tabular}} \\
\multicolumn{1}{c|}{} & \begin{tabular}[c]{@{}c@{}} \textit{\textbf{Tweets }}\\\textit{\textbf{count}} \end{tabular} & \begin{tabular}[c]{@{}c@{}} \textbf{\textit{Time }}\\\textbf{\textit{ period}} \end{tabular} &  &  &  &  &  &  &  &  \\ 
\arrayrulecolor{black}\cline{1-1}\arrayrulecolor{black}\cline{2-11}
 \textit{Jahanbin et al.}\cite{Jahanbin2020}\textit{}  & 364,080 & \begin{tabular}[c]{@{}c@{}} Dec. 2019 -\\Feb. 2020 \end{tabular} & ~ & ~ & X & ~ & ~ & ~ & ~ & ~ \\
 \textit{Banda et al.}\cite{banda2020largescale}]\textit{ }  & 30,990,645 & \begin{tabular}[c]{@{}c@{}} Jan.-\\Apr. 2020 \end{tabular} & X & ~ & ~ & ~ & ~ & ~ & ~ & ~ \\
 \textit{Medford et al.}\cite{Medford2020}\textit{}  & 126,049 & \begin{tabular}[c]{@{}c@{}} Jan.-\\Jan. 2020 \end{tabular} & X & X & ~ & X & X & ~ & ~ & ~ \\
 \textit{Singh et al.}\cite{Singh2020}\textit{}  & 2,792,513 & \begin{tabular}[c]{@{}c@{}} Jan. -\\Mar. 2020 \end{tabular} & X & X & X & X & ~ & ~ & ~ & ~ \\
 \textit{Lopez et al.}\cite{Lopez2020}\textit{}  & 6,468,526 & \begin{tabular}[c]{@{}c@{}} Jan.-\\Mar. 2020 \end{tabular} & X & X & X & ~ & ~ & ~ & ~ & ~ \\
 \textit{Cinelli et al.}\cite{Cinelli2020}\textit{}  & 1,187,482 & \begin{tabular}[c]{@{}c@{}} Jan. -\\Feb. 2020 \end{tabular} & ~ & X & ~ & X & ~ & ~ & ~ & ~ \\
 \textit{Kouzy et al.}\cite{Kouzy2020}\textit{}  & 673 & Feb 2020 & X & X & ~ & ~ & ~ & ~ & ~ & ~ \\
 \textit{Alshaabi et al.}\cite{Alshaabi2020}\textit{}  & Unknown & \begin{tabular}[c]{@{}c@{}} Mar. -\\Mar. 2020 \end{tabular} & X & X & ~ & ~ & ~ & ~ & ~ & ~ \\
 \textit{Sharma et al.}\cite{Sharma2020}\textit{}  & 30,800,000 & \begin{tabular}[c]{@{}c@{}} Mar. -\\Mar. 2020 \end{tabular} & X & X & X & X & X & ~ & ~ & X \\
 \textit{Chen et al.}\cite{Chen2020}\textit{}  & 8,919,411 & \begin{tabular}[c]{@{}c@{}} Mar. -\\Mar. 2020 \end{tabular} & X & ~ & ~ & ~ & ~ & ~ & ~ & ~ \\
 \textit{Schild et al.}\cite{Schild2020}\textit{}  & 222,212,841 & \begin{tabular}[c]{@{}c@{}} Nov. 2019 -\\Mar. 2020 \end{tabular} & X & X & ~ & X & ~ & ~ & ~ & X \\
 \textit{Yang et al.}\cite{Yang2020}\textit{}  & Unknown & \begin{tabular}[c]{@{}c@{}} Mar. -\\Mar. 2020 \end{tabular} & X & ~ & ~ & ~ & ~ & ~ & ~ & X \\
 \textit{Ordun et al.}\cite{Ordun2020}\textit{}  & 23,830,322 & \begin{tabular}[c]{@{}c@{}} Mar. -\\~Apr. 2020 \end{tabular} & X & X & ~ & X & ~ & ~ & ~ & X \\
 \textit{Yasin-Kabir et al.}\cite{Kabir2020}\textit{}  & 100,000,000 & \begin{tabular}[c]{@{}c@{}} March. -\\Apr. 2020 \end{tabular} & X & X & X & ~ & X & ~ & ~ & ~ \\
 \textit{\textbf{Ours}}  & \textbf{ 131,083,839} & \begin{tabular}[c]{@{}c@{}}\textbf{Jan. -~}\\\textbf{Jun. 2020}\end{tabular} & \textbf{ X } & \textbf{ X } & \textbf{ X } & \textbf{ X } & \textbf{ X } & \textbf{ X } & \textbf{ X } & \textbf{ X }
\end{tabular}
}
\end{table}

Table \ref{tab:papersComparison} provides a summary of most related studies in comparison to features studied in this research based on Twitter dataset.

\section{Datasets}
We collected two large-scale COVID-19 related datasets from Twitter and Instagram to study the topics discussed and people’s emotions on social networks during the pandemic. This section describes the collected datasets. 

\paragraph{Twitter} For our Twitter data collection, we collected 131,083,839 tweets posted between January 21, 2020, and June 19, 2020. We leveraged many tools, such as Hydrator \cite{Hydrator} and Twarc \cite{Twarc}, to rehydrate the Tweet IDs that are publicly available by Chen et al. \cite{Chen2020}. Moreover, we utilized twint \cite{Twint}, an advanced Twitter scraping \& OSINT tool in Python, to enrich our data collection. To concentrate on the pandemic-related tweets, we targeted tweets with hashtags, such as “\textit{\#covid19}”, “\textit{\#corona}”, “\textit{\#staysafe}”, “\textit{\#covid\_19}”, “\textit{\#covid2019}”, “\textit{\#lockdown}”, “\textit{\#stayhome}”, “\textit{\#quarantinelife}”, “\textit{\#coronacrisis}”, “\textit{\#coronavirus}”, “\textit{\#quarantine}”, etc. After filtering out retweets, i.e., focusing on unique tweets, the dataset was reduced by around 63\% (i.e., from 131,083,839 to 48,387,435 unique tweets). Table \ref{tab:TwitterDataset} shows some details about the collected tweets in terms of length and count.

\paragraph{Instagram} The Instagram posts are gathered using the open-source project called Instaloader \cite{Instaloader} and the publicly available Instagram Posts IDs provided by Zarei et al. \cite{Zarei2020}. Our collection of Instagram posts includes a total of 3,843 posts from January 5, 2020, to March 30, 2020. Concentrating on the pandemic-related posts, we collected posts with hashtags, such as “\textit{\#coronavirus}”, “\textit{\#covid19}”, “\textit{\#covid\_19}”, and “\textit{\#corona}”. We filtered out non-English posts and focused on analyzing posts written in English. The Instagram dataset contains 2,052 English posts in total. Table \ref{tab:InstagramDataset} shows some details about the collected Instagram dataset.

\begin{table}
\centering
\arrayrulecolor{black}
\captionsetup{justification=justified}
\captionof{table}{Twitter dataset information about tweets count, a tweet length in terms of characters (average, minimum, and maximum)}
\label{tab:TwitterDataset}
\resizebox{\linewidth}{!}{%
\begin{tabular}{l|cccccc|c}
\multicolumn{1}{c|}{\textbf{ ~ }} &  \textit{\textbf{January}}  &  \textit{\textbf{February}}  &  \textit{\textbf{March}}  &  \textit{\textbf{April}}  &  \textit{\textbf{May}}  &  \textit{\textbf{June}}  &  \textbf{\textit{Overall stats.} } \\ 
\arrayrulecolor{black}\cline{1-1}\arrayrulecolor{black}\cline{2-6}\arrayrulecolor{black}\cline{7-7}\arrayrulecolor{black}\cline{8-8}
 \textit{Total Tweets Count}  & 8,893,215 & 20,029,797 & 31,085,147 & 28,680,224 & 4,421,008 & 37,974,448 & 131,083,839 \\
 \textit{Unique Tweets Count}  & 2,842,466 & 3,946,405 & 13,160,649 & 15,331,109 & 4,222,937 & 8,883,869 & 48,387,435 \\
 \textit{Tweet Length (average number of characters)}  & 133.9 & 134.6 & 142 & 150.8 & 161.8 & 136.5 & 141.3 \\
 \textit{Tweet Length (minimum number of characters)}  & 4 & 2 & 3 & 3 & 3 & 2 & 2 \\
 \textit{Tweet Length (maximum number of characters)}  & 1,012 & 1,013 & 1,023 & 1,021 & 1,014 & 1,007 & 1,023
\end{tabular}
}
\end{table}


\begin{table}
\centering
\arrayrulecolor{black}
\captionsetup{justification=justified}
\captionof{table}{Instagram dataset information about posts count, a post length in terms of characters (average,minimum, and maximum)}
\label{tab:InstagramDataset}
\resizebox{\linewidth}{!}{%
\begin{tabular}{>{\hspace{0pt}}m{0.42\linewidth}>{\hspace{0pt}}m{0.027\linewidth}|>{\centering\hspace{0pt}}m{0.086\linewidth}>{\hspace{0pt}}m{0.027\linewidth}>{\centering\hspace{0pt}}m{0.094\linewidth}>{\hspace{0pt}}m{0.027\linewidth}>{\centering\hspace{0pt}}m{0.065\linewidth}>{\hspace{0pt}}m{0.027\linewidth}|>{\centering\hspace{0pt}}m{0.125\linewidth}>{\hspace{0pt}}m{0.027\linewidth}}
~ &  & \textit{\textbf{January}} &  & \textit{\textbf{February}} &  & \textit{\textbf{March}} &  & \textit{\textbf{Overall stats}.} &  \\ 
\hline
\textit{Total Posts Count} &  & 749 &  & 1,530 &  & 1,564 &  & 3,843 &  \\
\textit{Post Length (average number of characters)} &  & 392 &  & 372 &  & 402 &  & 388.1 &  \\
\textit{Post Length (minimum number of characters)} &  & 4 &  & 2 &  & 2 &  & 2 &  \\
\textit{Post Length (maximum number of characters)} &  & 2,199 &  & 2,376 &  & 2,228 &  & 2,376 & 
\end{tabular}
}
\end{table}

\section{Sentiment Analysis} 

Although several studies have conducted sentiment analysis on social network data \cite{Medford2020,Sharma2020,Kabir2020}, the main purpose of this research is to conduct a comprehensive analysis to understand people’s worldwide reaction to the pandemic. Therefore, this study aims to incorporate sentiment analysis to gain insights about people’s reactions to topics and trends in different time periods so it becomes feasible to discover geo-temporal patterns of users’ behavior. This section describes the methods and results of the sentiment analysis on our datasets.
 \newline

\begin{figure}[h]
    \centering
    \captionsetup{justification=justified}
    \includegraphics[width=\textwidth]{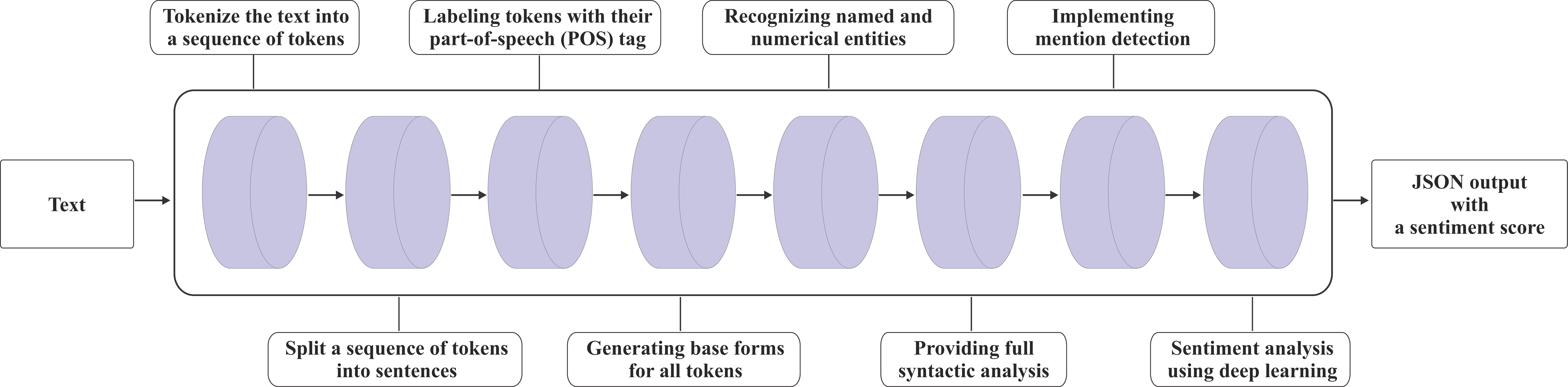}
    \caption{The overall architecture of Stanford CoreNLP to find sentiment classes of Twitter and Instagram datasets.}
    \label{fig:STCoreNLP}
\end{figure}

\paragraph{Methods} Before conducting our analysis on the collected datasets, we adopted a pre-processing stage which includes cleaning the data, removing URL links, emails, user mentions, punctuations, stopwords, and converting emojis and emoticons into words. As deep learning techniques has shown powerful performance in intelligent processing of data in several domain \cite{ABUHMED2021106688,9144577,9007368,abuhamad2020multi}, we leveraged Stanford deep learning based NLP model called CoreNLP for the sentiment analysis \cite{Manning2014}. CoreNLP toolkit provides an extensible pipeline supporting core NLP tasks, and it is based on a compositional model over binarized trees of sentences using deep learning \cite{Fischer2004}, and it contains a collection of pre-trained models. Due to the Figure \ref{fig:STCoreNLP} illustrates the CoreNLP architecture, starting from providing an input content and ending with an output of all the analysis information as a JSON file. The output includes the sentiment score represented with a value ranging from zero to four, where four means that the input is very positive and zero means the input is very negative. 

\begin{figure}[h]
    \centering
    \captionsetup{justification=justified}
\begin{tabular}{cc}
    {\includegraphics[width=0.46\textwidth]{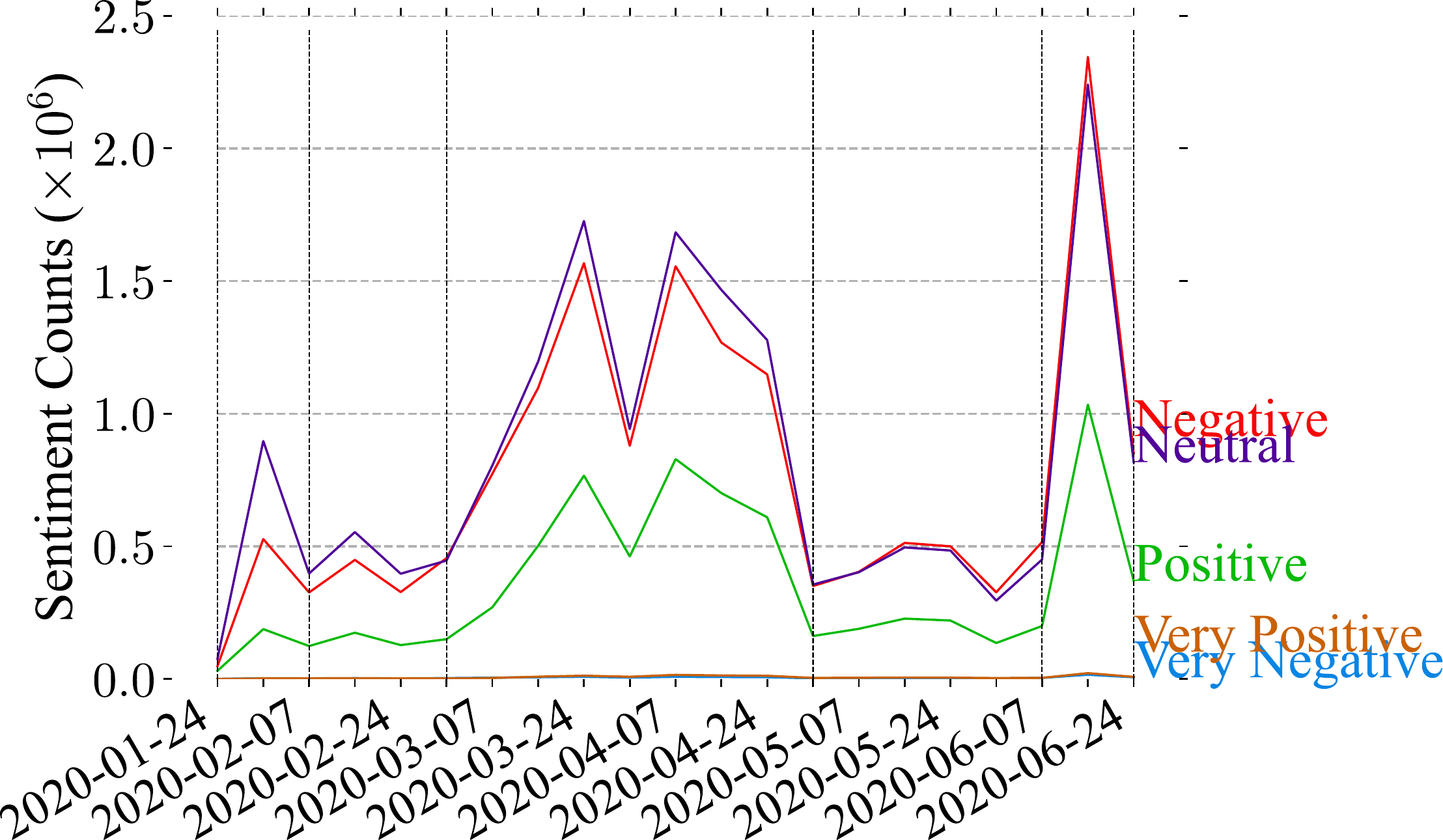}} & {\includegraphics[width=0.46\textwidth]{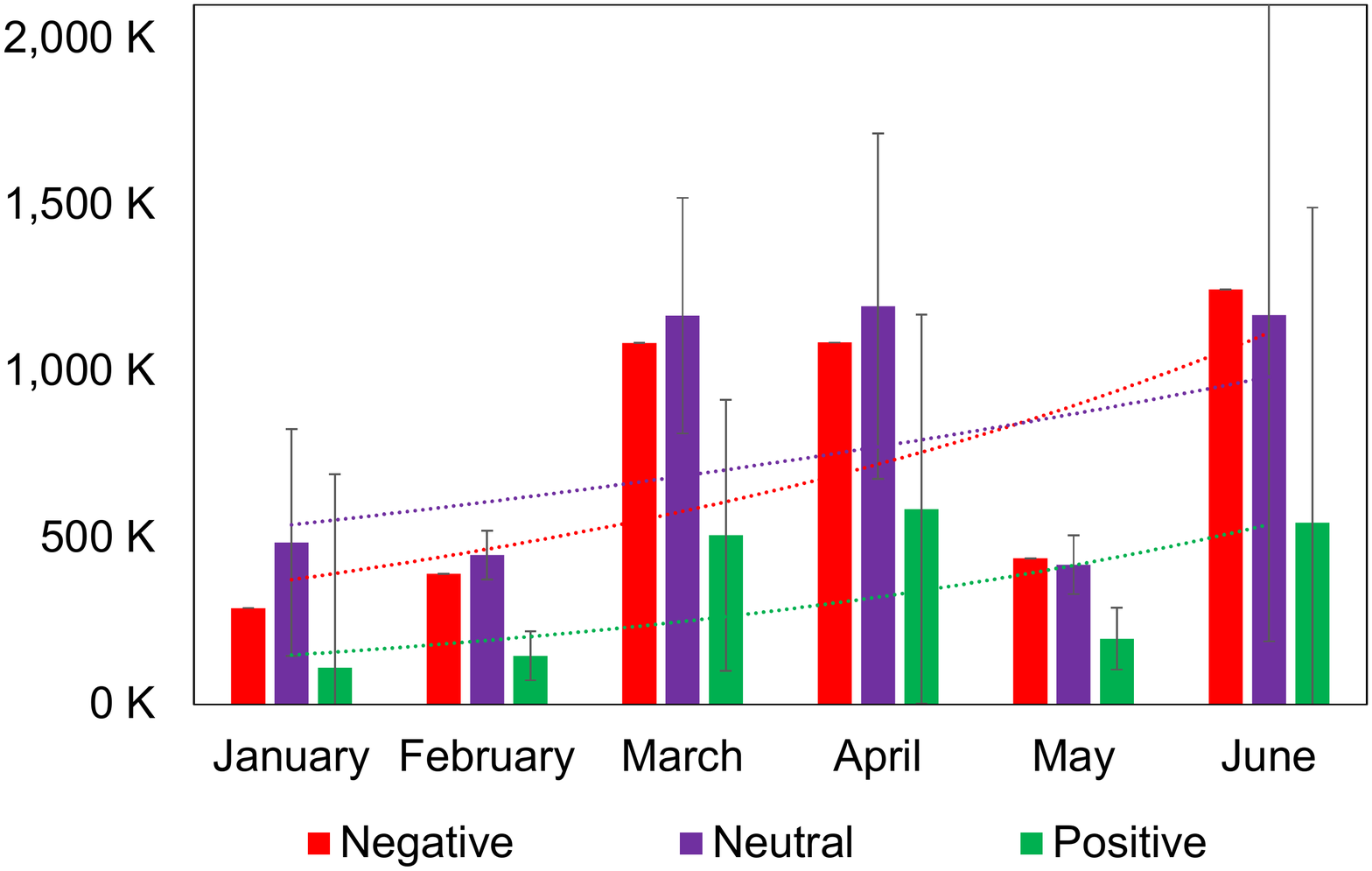}} \\ \multicolumn{2}{c}{(a)} \\[5pt]
    {\includegraphics[width=0.46\textwidth]{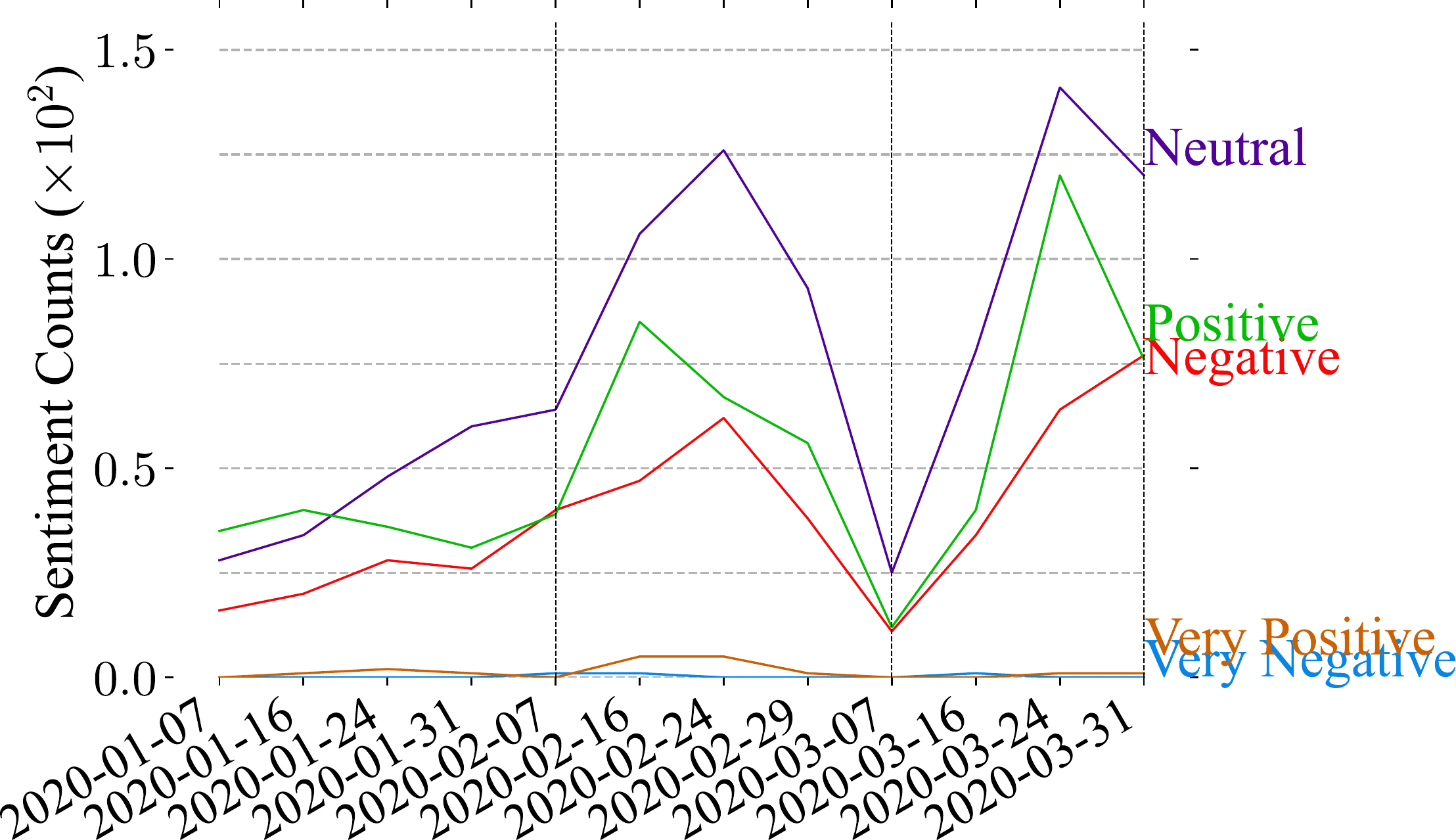}} & {\includegraphics[width=0.46\textwidth]{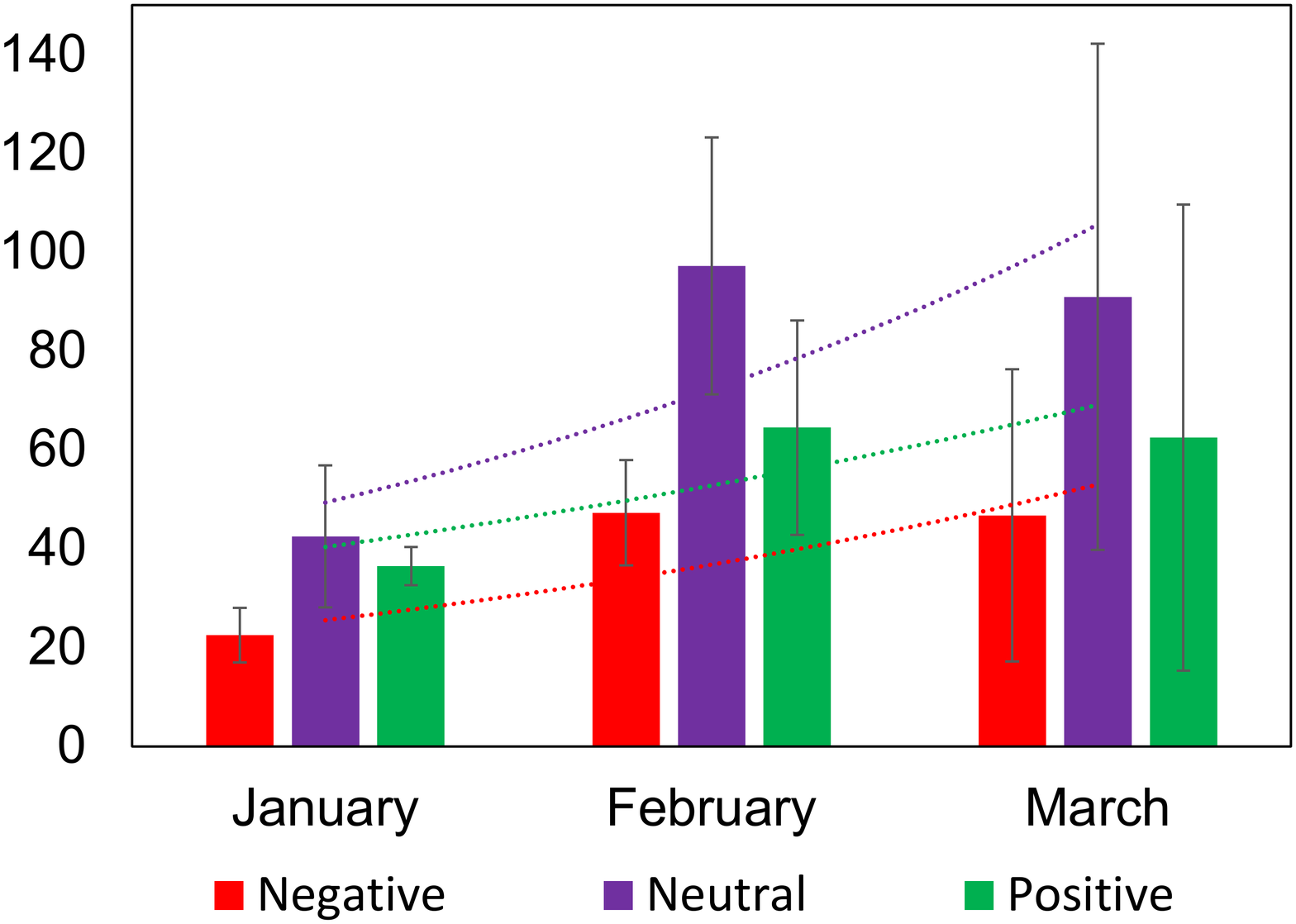}} \\ \multicolumn{2}{c}{(b)} \\[5pt]
\end{tabular} 
    \caption{Weekly and monthly sentiment analysis of Twitter (a) and Instagram (b) datasets based on five sentiment classes (very negative, negative, neutral, positive, very positive). For weekly analysis (right side), spikes are highlighted with vertical black dashed lines, and trends are illustrated in the monthly analysis (left side) after merging the very positive and very negative to their corresponding classes.}
    \label{fig:WeeklySentiment}
\end{figure}

\paragraph{Twitter} We performed the analysis on all tweets based on the five sentiment categories (very negative, negative, neutral, positive, and very positive). Figure \ref{fig:WeeklySentiment} (a) represents the distribution of sentiments in all the collected tweets between January 24, 2020, and June 24, 2020.  Observing the sentiment during the data collection timeline, Figure \ref{fig:WeeklySentiment} shows that most of the tweets are neutral in general. However, most tweets in June are negative.  In March, April, and June, there are significant spikes that occurred, as well as we can consider the rise in January as a spike because of a small but marked increase. These spikes are highlighted in the figure with vertical yellow dashed lines to indicate which part of the dataset is used to identify topics discussed on each sentiment type in the next section. 

\paragraph{Instagram} The same sentiment categories (very negative, negative, neutral, positive, very positive) are applied to the Instagram dataset. Figure \ref{fig:WeeklySentiment} (b) shows the sentiment analysis of collected posts between the period of January 7, 2020, and March 31, 2020. In the figure, it is shown that the neutral category is dominant over others. In contrast with the results of Twitter, the second dominant category is positiveness while it is the negative type in Figure \ref{fig:WeeklySentiment} (a). The result has two spikes in February and March posts (highlighted with vertical black dashed lines).

\section{Topic Modeling on Sentiment Types}

In this section, we conduct the topic modeling task on data collected during spikes of users’ reactions, i.e., covering different sentiment types (negative, positive, neutral). As observed by our sentiment analysis, we re-categorized sentiments into three types, instead of five, to discover topics discussed on the social media platforms that attracted these sentiments. This per-sentiment topic modeling enables the understanding of people’s reactions to certain topics.

\begin{figure}[h]
    \centering
    \captionsetup{justification=justified}
    \includegraphics[width=0.8\textwidth]{ 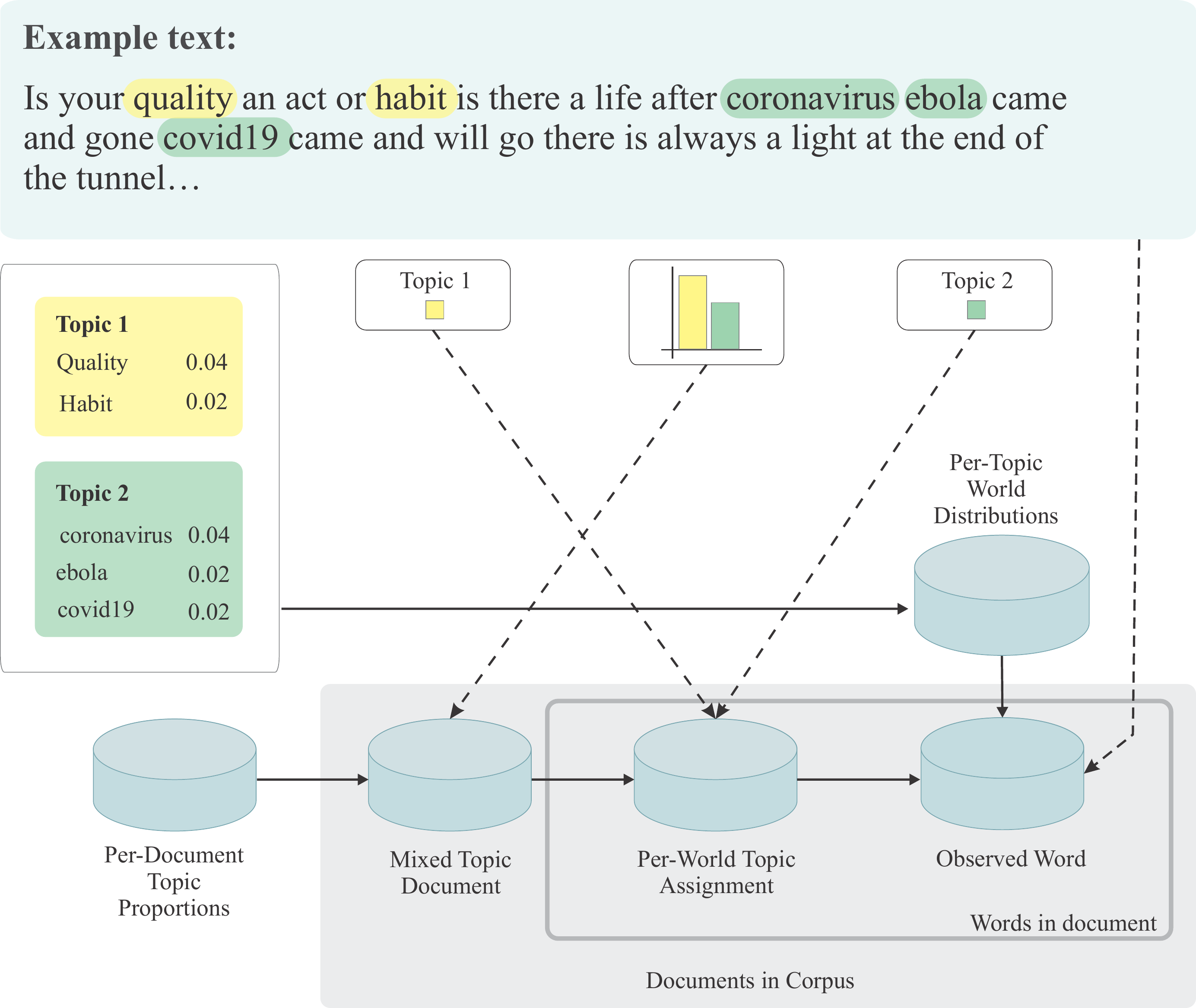}
    \caption{Plate notation of Latent Dirichlet Allocation (LDA) to discover topics discussed on both social networks}
    \label{fig:LDA}
\end{figure}

\paragraph{Methods} For topic modeling, we utilized the popular Latent Dirichlet Allocation (LDA) \cite{rehurek_lrec} method. We leverage the implementation of LDA in the Gensim package along with the Mallet \cite{Mallet} implementation to extract the topics discussed during the spike periods of various sentiment types. LDA with Mallet is an efficient technique to extract topics with sufficient topic segregation. Figure \ref{fig:LDA} shows a brief description of the LDA model. As shown in Figure \ref{fig:LDA}, the two corpus-level parameters (documents in corpus and words in the document) represent a Bayesian method for randomly sampling a mixture of topics for each document. Firstly, the LDA process goes through each document and assigns each word in each document (tweet or post) to one of the N topics (N is set by us) randomly. Then LDA checks each word in all documents again and calculates the proportion of words in each document and the proportion of assigning words to each topic. By doing this process repeatedly, LDA shifts words around topics to find the most suitable one. At the end of the process, good quality topics that are clear, meaningful, and segregated are extracted.

To improve the generation of topics, we have taken into consideration some key factors, such as the quality of the presented text, the number of topics per tweet or post, the total number of topics, the various parameters of the algorithm. For the quality of the presented text, we adopt a preprocessing stage to make the input content ready for the analysis. This preprocessing stage includes the removal of stopwords, emails, user mentions, extra spaces, URL links, and punctuations. It also includes tokenization, lemmatization, and the extraction of unigrams, bigrams, and trigrams.

\paragraph{LDA Model Selection} We used the topic coherence score to estimate how well a model provides different topics. To this end, we experimented with a different number of topics, from 2 to 30 topics, to evaluate the LDA model performance generating topics with a high coherence value. For example, Ordun et al. \cite{Ordun2020} selected the LDA model that generated 20 topics with a 0.344 coherence score. Similarly, Sharma et al. \cite{Sharma2020} identified 20 different topics using topic modeling that is based on compressed text classification.

\begin{table}
\centering
\small
\arrayrulecolor{black}
\captionsetup{justification=justified}
\captionof{table}{Total number of topics of spike periods per sentiment type and general topics’ contribution per generated topics}
\label{tab:SpikeTable}
\resizebox{\linewidth}{!}{%
\begin{tabular}{>{\hspace{0pt}}m{0.211\linewidth}>{\centering\hspace{0pt}}m{0.027\linewidth}|>{\centering\hspace{0pt}}m{0.125\linewidth}>{\centering\hspace{0pt}}m{0.106\linewidth}>{\centering\hspace{0pt}}m{0.115\linewidth}|>{\centering\hspace{0pt}}m{0.125\linewidth}>{\centering\hspace{0pt}}m{0.106\linewidth}>{\centering\arraybackslash\hspace{0pt}}m{0.115\linewidth}}
\multicolumn{2}{>{\hspace{0pt}}m{0.28\linewidth}!{\color{black}\vrule}}{\multirow{3}{0.7\linewidth}{\hspace{0pt}\textbf{General topics}\textit{}}} & \multicolumn{3}{>{\centering\hspace{0pt}}m{0.346\linewidth}!{\color{black}\vrule}}{\textit{\textbf{Twitter}}} & \multicolumn{3}{>{\centering\arraybackslash\hspace{0pt}}m{0.346\linewidth}}{\textit{\textbf{Instagram}}} \\ 
\cline{3-5}\arrayrulecolor{black}\cline{6-8}
\multicolumn{2}{>{\hspace{0pt}}m{0.238\linewidth}!{\color{black}\vrule}}{} & \multicolumn{3}{>{\centering\hspace{0pt}}m{0.346\linewidth}|}{\textit{\textbf{Sentiment types}}} & \multicolumn{3}{>{\centering\arraybackslash\hspace{0pt}}m{0.346\linewidth}}{\textit{\textbf{Sentiment types}}} \\ 
\cline{3-5}\arrayrulecolor{black}\cline{6-8}
\multicolumn{2}{>{\hspace{0pt}}m{0.238\linewidth}!{\color{black}\vrule}}{} & \textit{\textbf{Negative}} & \textit{\textbf{Neutral}} & \textit{\textbf{Positive}} & \textit{\textbf{Negative}} & \textit{\textbf{Neutral}} & \textit{\textbf{Positive}} \\ 
\hline
\textit{\textbf{Economy}} & ~ & 1 & 1 & 2 & 0 & 0 & 0 \\
\textit{\textbf{Health}} & ~ & 4 & 2 & 1 & 2 & 1 & 1 \\
\textit{\textbf{Social}} & ~ & 5 & 3 & 6 & 5 & 1 & 4 \\
\textit{\textbf{Politics}} & ~ & 1 & 1 & 1 & 0 & 0 & 0 \\
\textit{\textbf{Tourism}} & ~ & 1 & 1 & 0 & 1 & 0 & 0 \\
\textit{\textbf{Fashion}} & ~ & 0 & 0 & 0 & 0 & 0 & 1 \\ 
\cline{1-2}\arrayrulecolor{black}\cline{3-4}\arrayrulecolor{black}\cline{5-5}\arrayrulecolor{black}\cline{6-8}
\textit{\textbf{Topics Count}} & ~ & 12 & 8 & 10 & 8 & 2 & 6 \\ 
\arrayrulecolor{black}\cline{1-2}\arrayrulecolor{black}\cline{3-4}\arrayrulecolor{black}\cline{5-5}\arrayrulecolor{black}\cline{6-8}
\textit{\textbf{Coherent Score}} & ~ & 0.4754 & 0.4132 & 0.4533 & 0.4810 & 0.5734 & 0.4175
\end{tabular}
}
\arrayrulecolor{black}
\end{table}

\paragraph{\textbf{Remarks}} We only apply topic modeling on Twitter and Instagram data during the spike periods shown in Figure \ref{fig:WeeklySentiment}. Since the size of the data in those periods is different, we have different numbers of topics per spike period. The generated topics are then assigned into a set of predefined general categories, namely, Economy, Health, Social, Politics, and Tourism. Table \ref{tab:SpikeTable} shows the results of our topic modeling technique. In the table, total topics with corresponding coherent scores, as well as the number of topics that belong to general categories, are given. It should be noted that a new category called “Fashion” is added to the table as our analysis on the Instagram dataset generated this topic.

\begin{figure}[h]
    \centering
    \captionsetup{justification=justified}
    \includegraphics[width=\textwidth]{ 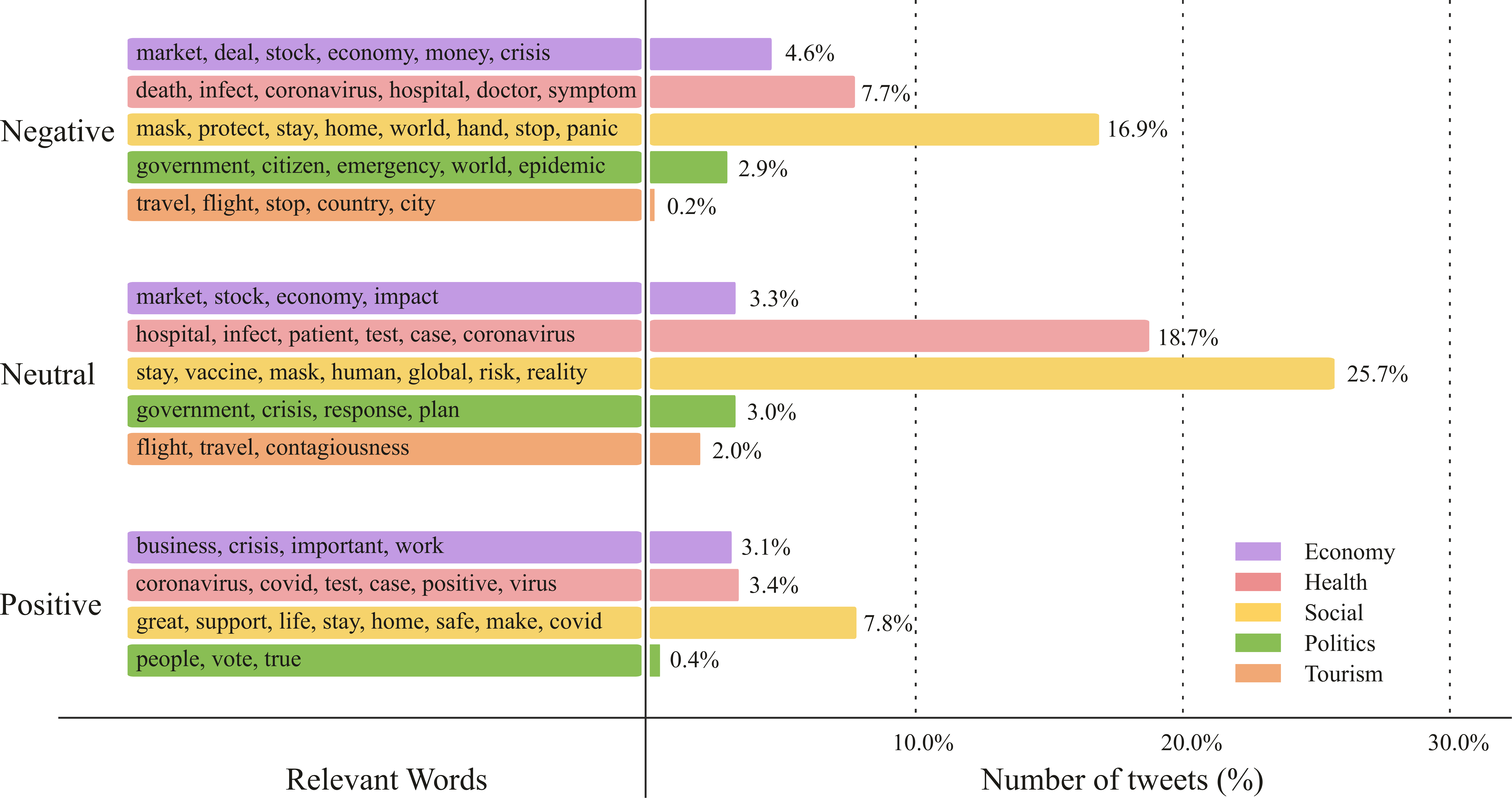}
    \caption{Topics generated for each sentiment type and number of tweets belonged to each topic (Twitter)}
    \label{fig:TwitterSentiment}
\end{figure}

\paragraph{Twitter} Figure \ref{fig:TwitterSentiment} shows the result of topic modeling on Twitter dataset. The figure is divided into two parts by the vertical line: the left part shows corresponding keywords that express topics, and the right part shows the distribution of each topic on the analyzed dataset (in percentage), as well as, the topics are grouped by three sentiment classes: negative, positive, and neutral. It is interesting to note that the rank distribution of topics is the same for all sentiment types. The figure shows that tweets about social-related topics are the most common among other categories, with about 50.4\% of the total tweets. The most observed keywords of various topics describe the people’s intent to notify and advise others with quarantine rules and ways to stay safe.
On the other hand, the least discussed topic is about travel-related tweets, as shown for the negative and neutral groups with 0.2\% and 2\%, respectively. For the positive category, the least discussed topic is the politics category, with 0.4\% of the total tweets. Our analysis shows that most tweets belong to the neutral type populating proportionally the topics observed by the topic modeling task. Tweets assigned to the economy-related topic attract more negative sentiments with 4.6\% of the total tweets compared to 3.3\% and 3.1\% on the neutral and positive sentiments, respectively. This trend of people’s reaction towards economy-related topics reflects their concerns about the impact of COVID-19 on the economy.

\begin{figure}[h]
    \centering
    \captionsetup{justification=justified}
    \includegraphics[width=\textwidth]{ 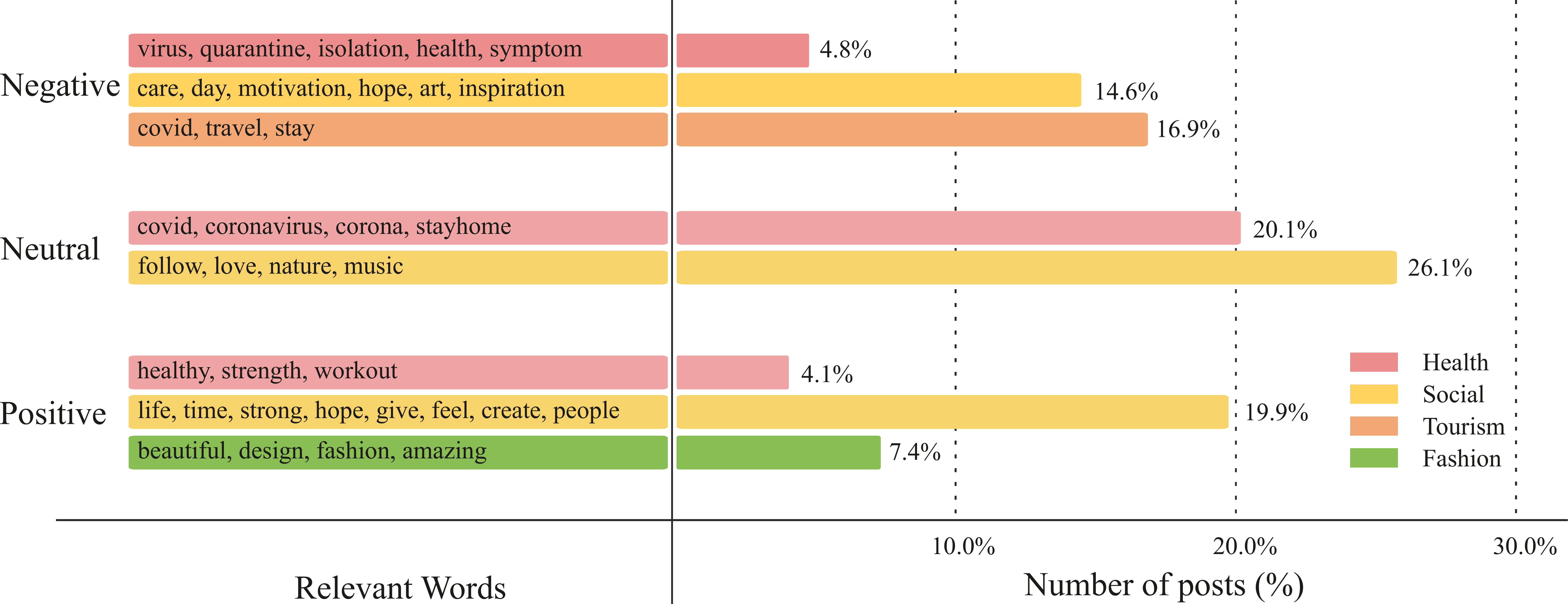}
    \caption{Topics generated for each sentiment type and number of tweets belonged to each topic (Instagram)}
    \label{fig:InstagramSentiment}
\end{figure}

\paragraph{Instagram} Figure \ref{fig:InstagramSentiment} shows the distributions of topics from the Instagram dataset are displayed in, categorized by the various sentiments. Compared to those observed from the Twitter dataset, there are fewer topics debated since users of Instagram publish fashion-related positive posts with a total of 7.4\% of the entire dataset. Social-related topics are the common topics in this platform, making up 14.6\% of the negative posts, 26.1\% of the neutral posts, and 19.9\% of the positive posts. The relevant words of the social-related topic highlight users inspirational thoughts to stay strong in the fight against COVID-19. For example, one post states the following: “\textit{is your quality an act or habit is there a life after coronavirus ebola came and gone covid19 came and will go there is always a light at the end of the tunnel…}”. Observing the topics on Instagram, the health-related topic is the second main topic in all sentiment.

\section{Geo-Temporal Analysis}

We started our geo-temporal analysis by identifying countries’ mentions on posts from Instagram and Twitter to explore the geo-temporal patterns in our data collection. Afterward, we investigate people’s reactions to topics discussed in those countries. Our study provides answers to the following questions: (1) what countries are mainly attached to the pandemic posts on social networks? (2) what are the top topics connected to the top-mentioned countries during the pandemic? (3) what are the geo-temporal trends and patterns observed across topics and localities during the pandemic on social networks?

\subsection{Country-based Analysis}

This section provides analysis and methods of tracking countries in posts on social media platforms. This process aims to define the top-mentioned countries in our data collection for further geo-temporal analysis.

\paragraph{Methods} To analyze and track countries mentioned on Twitter and Instagram, we utilize the NER technique implemented by Stanford CoreNLP \cite{Manning2014}. Stanford NER tagger assigns named entities in texts to recognize people, values, dates, places, and organizations. Using the Stanford NER tagger, we determine country mentions on all tweets and posts in our datasets. Afterward, we sort countries in descending order based on the number of mentions to select the top-5 mentioned countries in our data collection.

\begin{figure}[h]
    \centering
    \captionsetup{justification=justified}
    \includegraphics[width=0.9\textwidth]{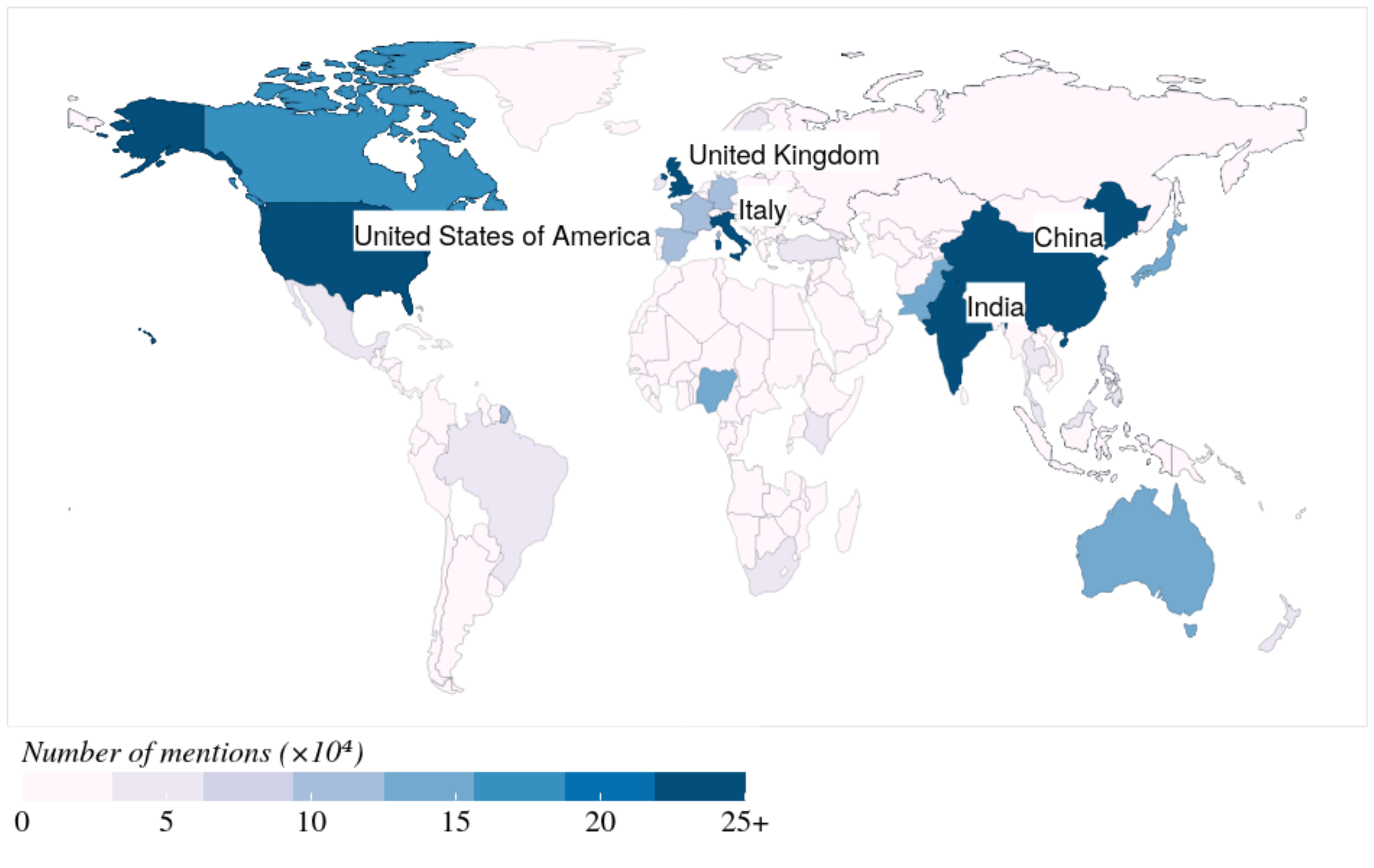}
    \caption{Countries that are mentioned in tweets}
    \label{fig:TwitterCountries}
\end{figure}

\paragraph{Twitter} We note that the adopted NER technique identifies not only country names but also cities and other places in a given input. However, we focus only on the mentions of countries to select the top mentioned countries. Based on our analysis, our dataset includes mentions of 180 countries, and China is the most mentioned country with 3,035,389 mentions. The next four top mentioned countries are the USA (695,131), India (541,851), the UK (521,290), and Italy (283,096). Figure \ref{fig:TwitterCountries} depicts the results of the number of mentions of countries in our dataset.

\begin{figure}[h]
    \centering
    \captionsetup{justification=justified}
    \includegraphics[width=0.9\textwidth]{ 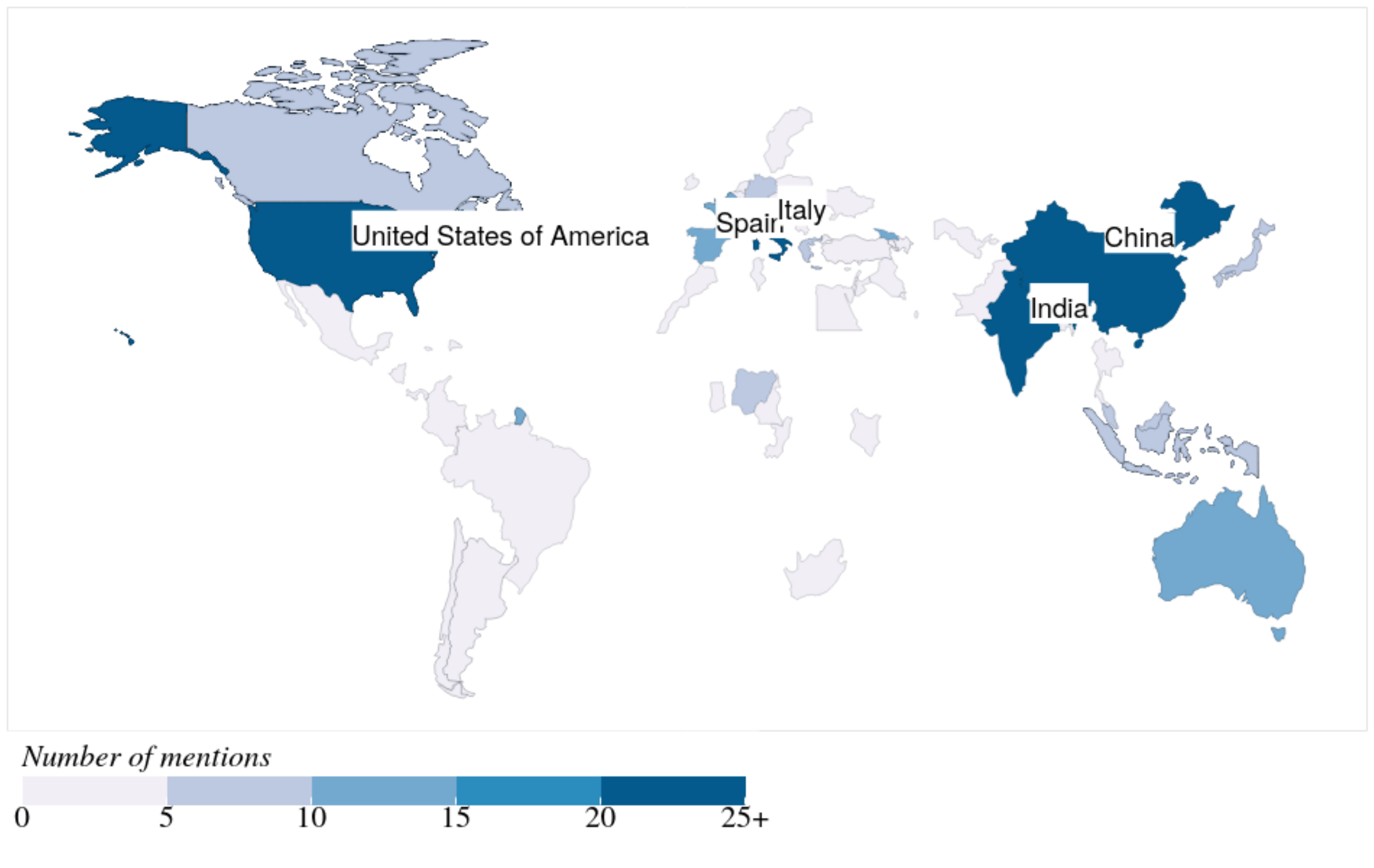}
    \caption{Countries that are mentioned in Instagram posts}
    \label{fig:InstagramCountries}
\end{figure}

\paragraph{Instagram} Using the NER technique on Instagram posts, we extracted 61 countries, where the top-mentioned countries are China, India, the USA, Italy, and Spain as demonstrated in Figure \ref{fig:InstagramCountries}. Figures \ref{fig:TwitterCountries} and \ref{fig:InstagramCountries} show that the top-mentioned countries on both social networks are almost the same, except for Spain and the UK. This supports the fact that these are the key countries in which a high number of coronavirus infections were reported within the corresponding periods \cite{WorldHealthOrganization}. To gain insights into people’s reactions and attitudes towards topics discussed in these countries, we conducted a detailed analysis of the top-mentioned countries including: sentiment analysis, topic modeling, word2vec-based analysis, and locality analysis.

\subsection{Sentiment Analysis of Top Countries}

This section provides sentiment analysis on data records observed with the top-5 mentioned countries in our data collection. This analysis enables the understanding of attitudes and reactions towards topics related to these countries during the pandemic.

\paragraph{Twitter} Figure \ref{fig:TwitterCountriesSentiment} illustrates the sentiment analysis of the top-5 countries, i.e., China, USA, India, UK, and Italy, from the Twitter dataset. The negative reactions, sentiment, observed in tweets mentioning the top-5 countries are more obvious than other sentiments over time. Figure \ref{fig:TwitterCountriesSentiment} (a) shows that China started with having more neutral tweets in January 2020. Interestingly, all countries except for China experienced an improvement in the growth of tweets in March, as shown in Figure \ref{fig:TwitterCountriesSentiment}. The figure also shows that the USA, India, and the UK charts are similar with respect to sentimentality spikes in March, April, and June, and they reached their highest points in June while Italy experienced its peak between March 16, 2020, and March 31, 2020.

\begin{longtable}{cc}
    {\includegraphics[width=0.5\linewidth]{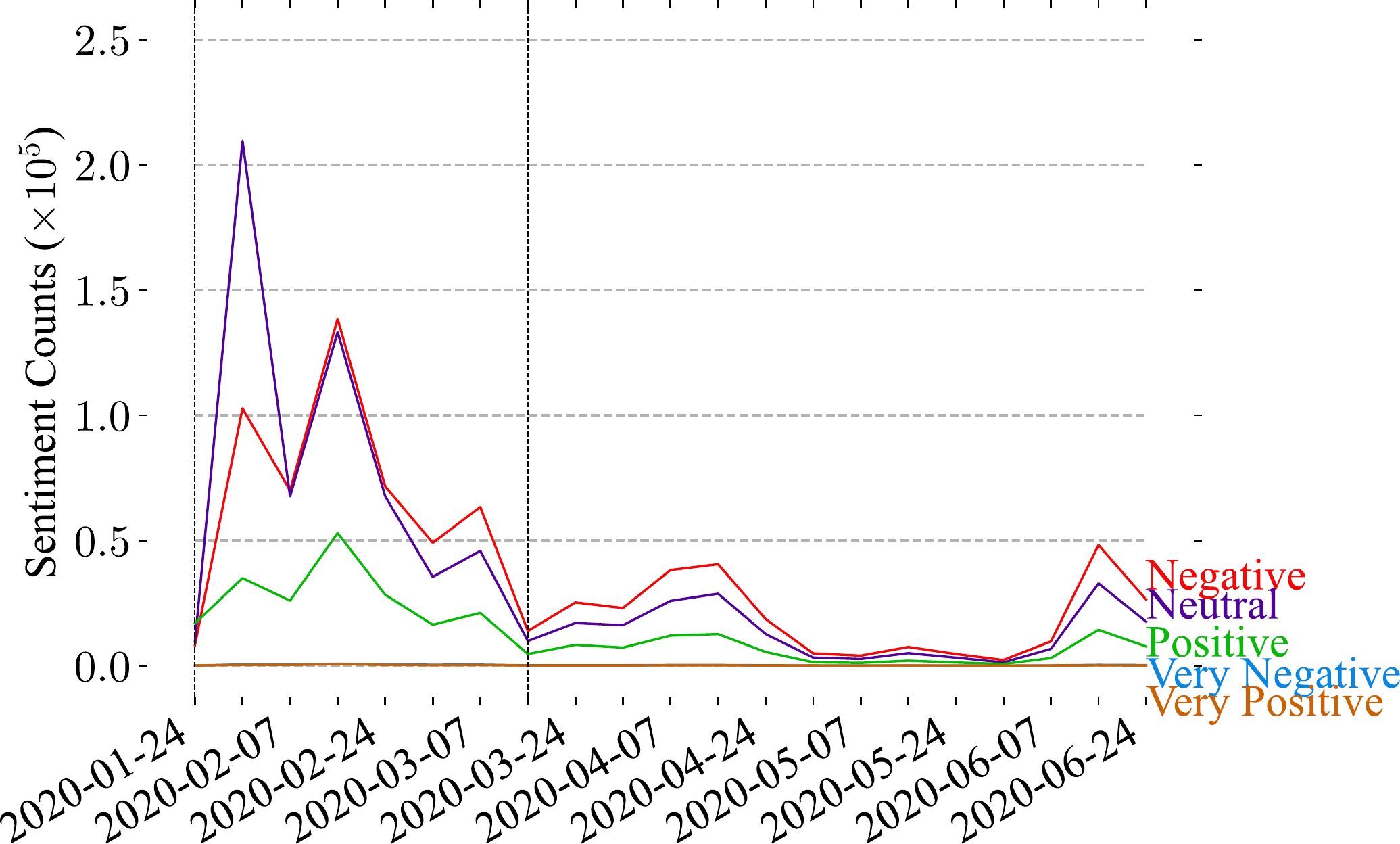}} & {\includegraphics[width=0.5\linewidth]{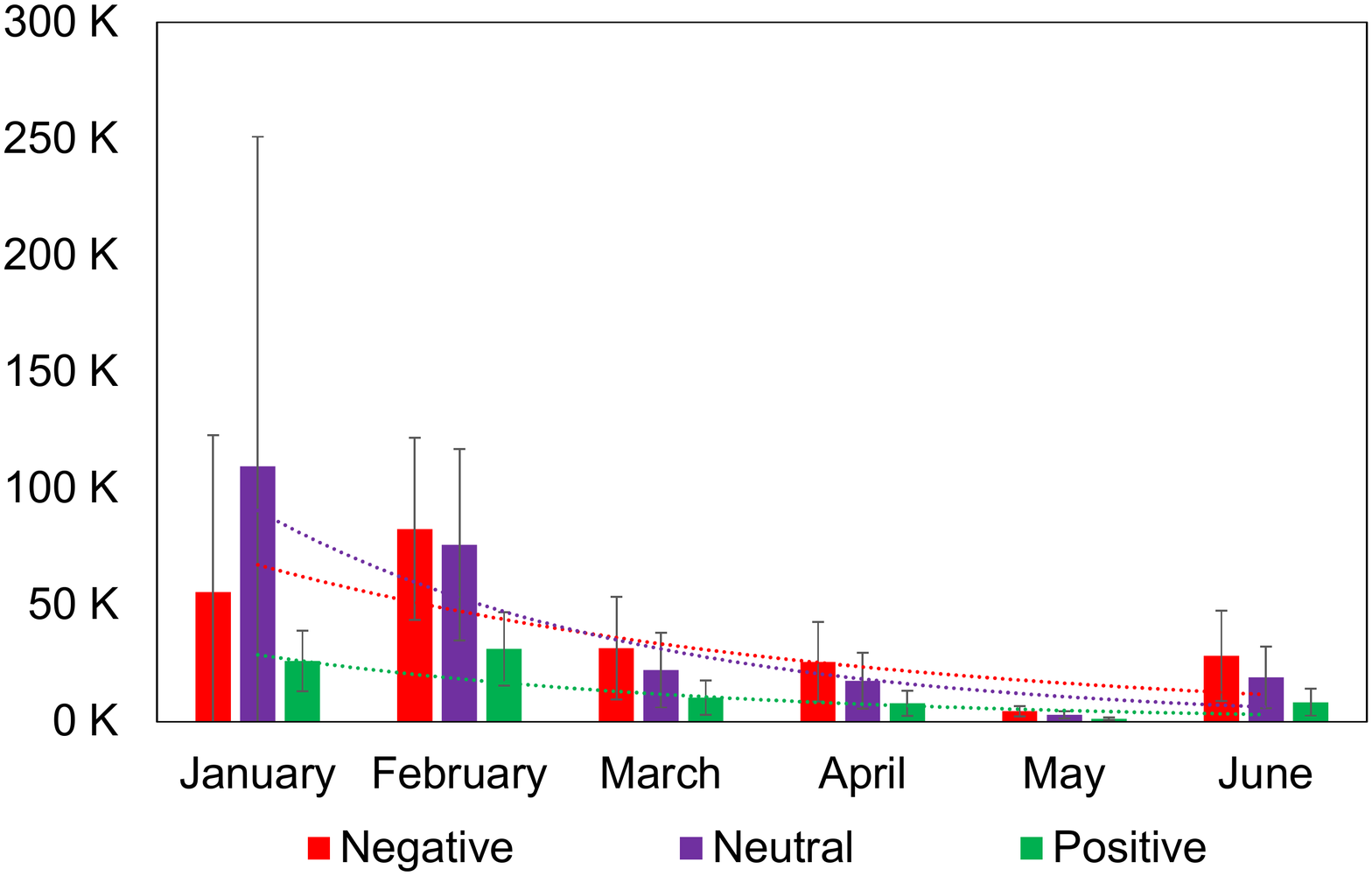}} \\ \multicolumn{2}{c}{(a) China} \\
    {\includegraphics[width=0.5\linewidth]{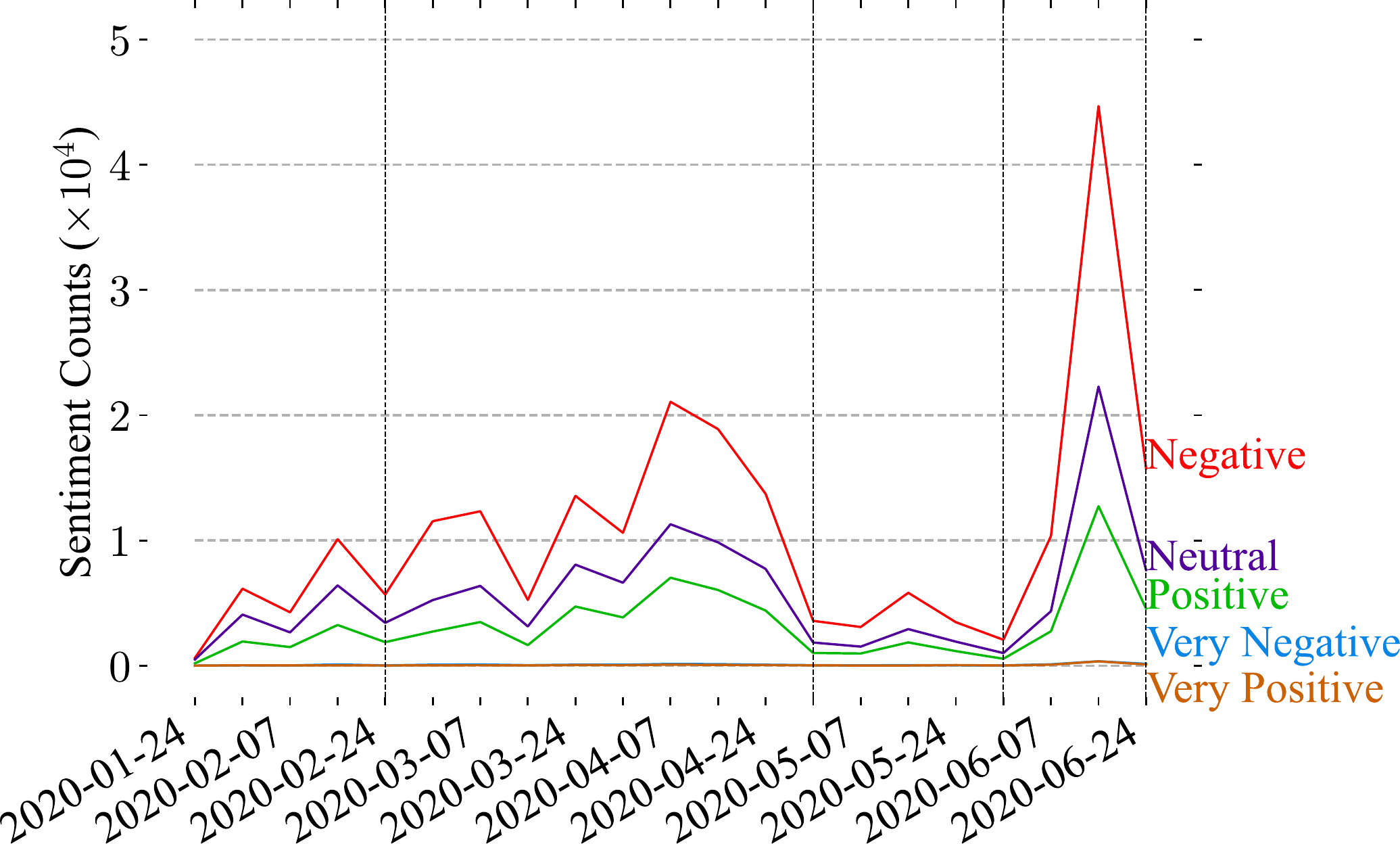}} & {\includegraphics[width=0.5\linewidth]{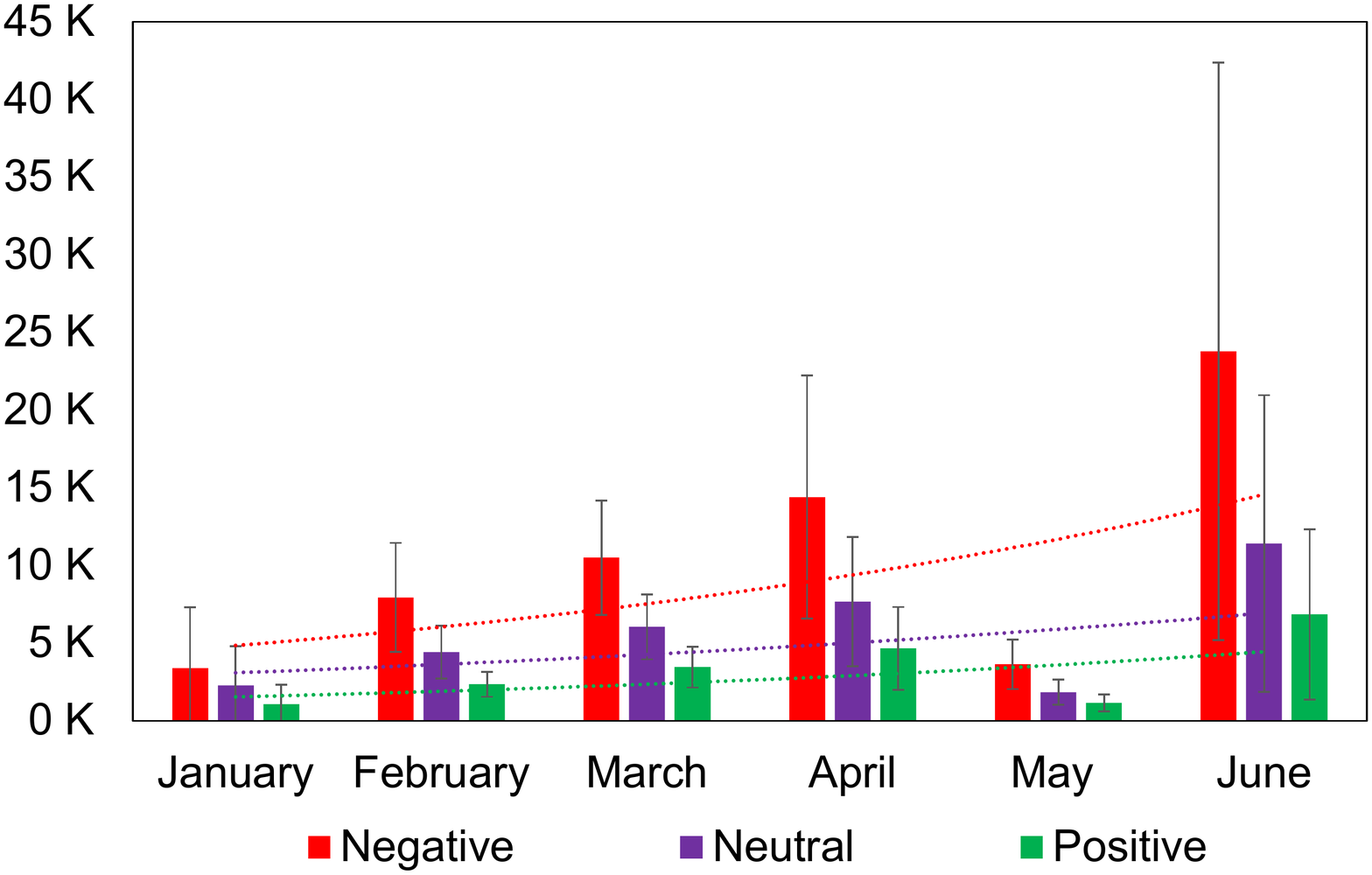}} \\ \multicolumn{2}{c}{(b) USA} \\
    {\includegraphics[width=0.5\linewidth]{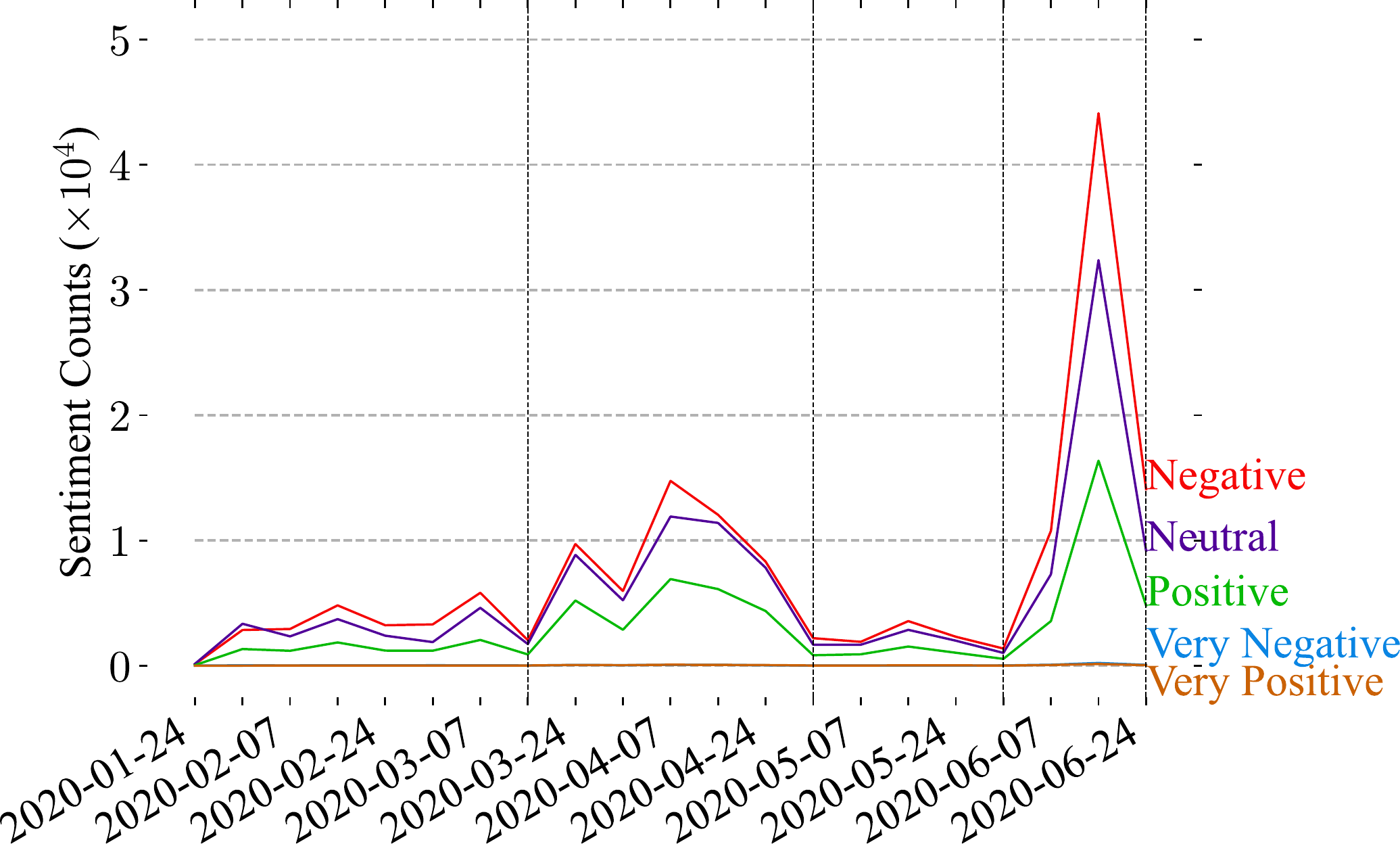}} & {\includegraphics[width=0.5\linewidth]{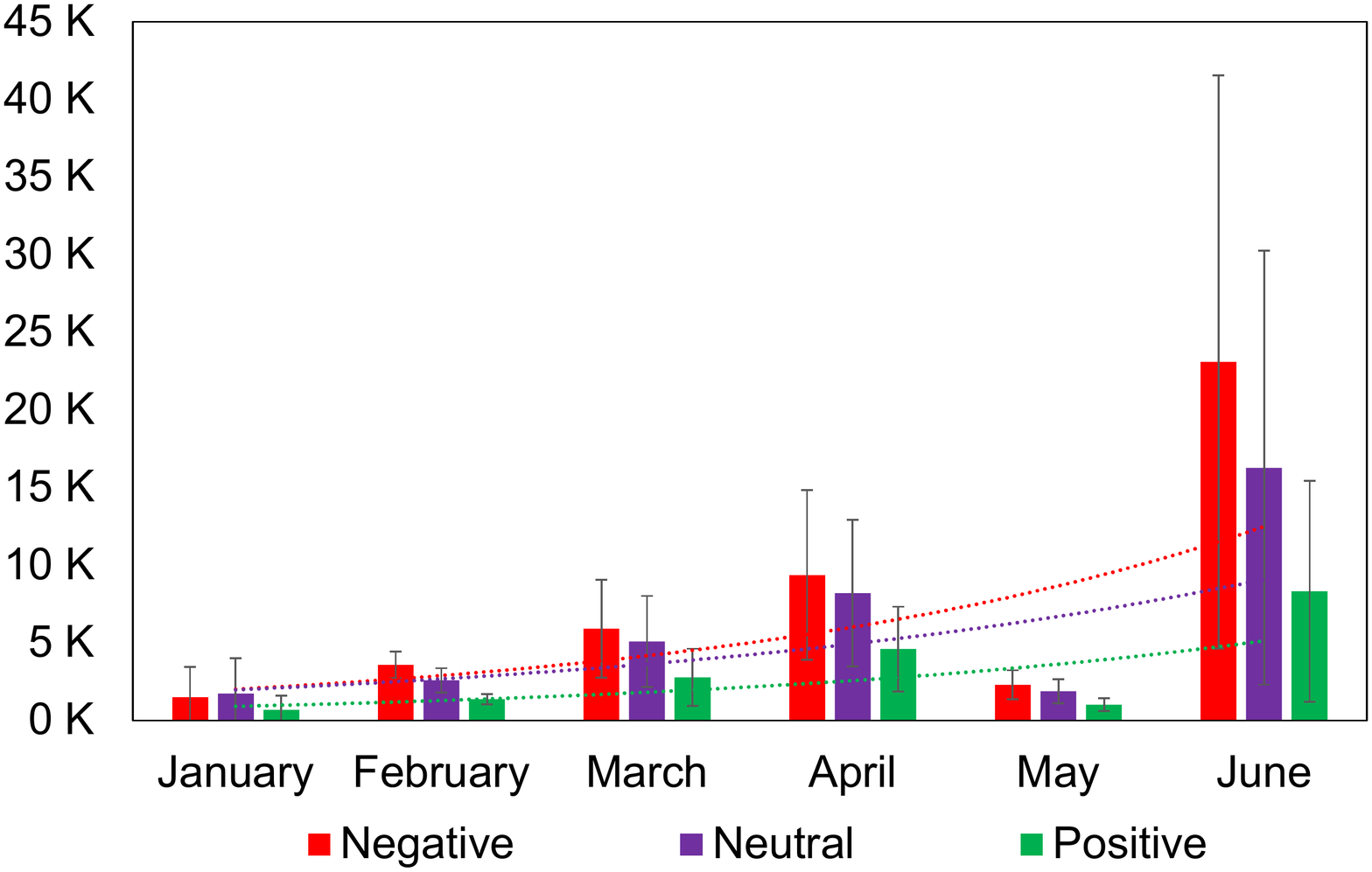}} \\ \multicolumn{2}{c}{(c) India} \\
    {\includegraphics[width=0.5\linewidth]{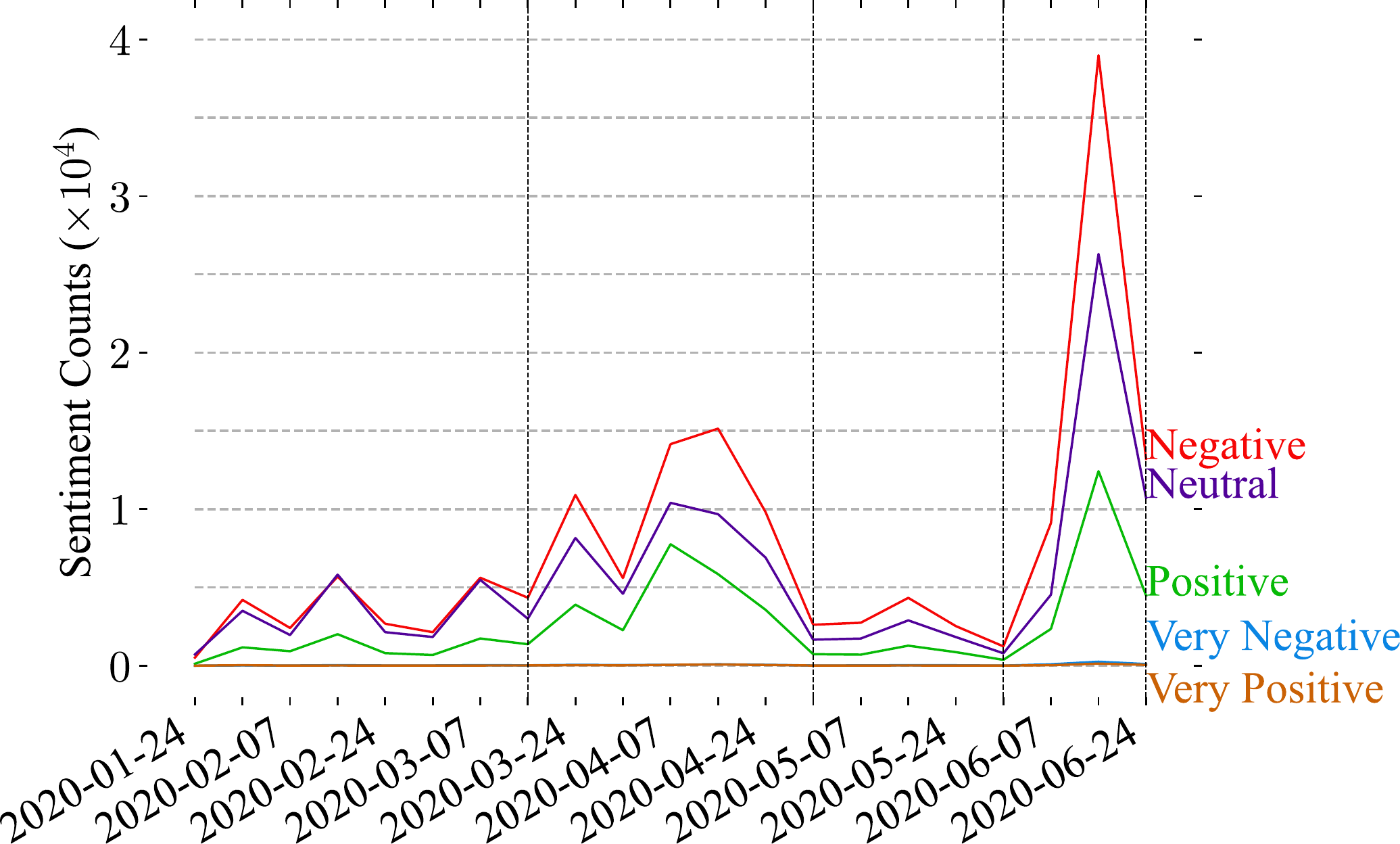}} & {\includegraphics[width=0.5\linewidth]{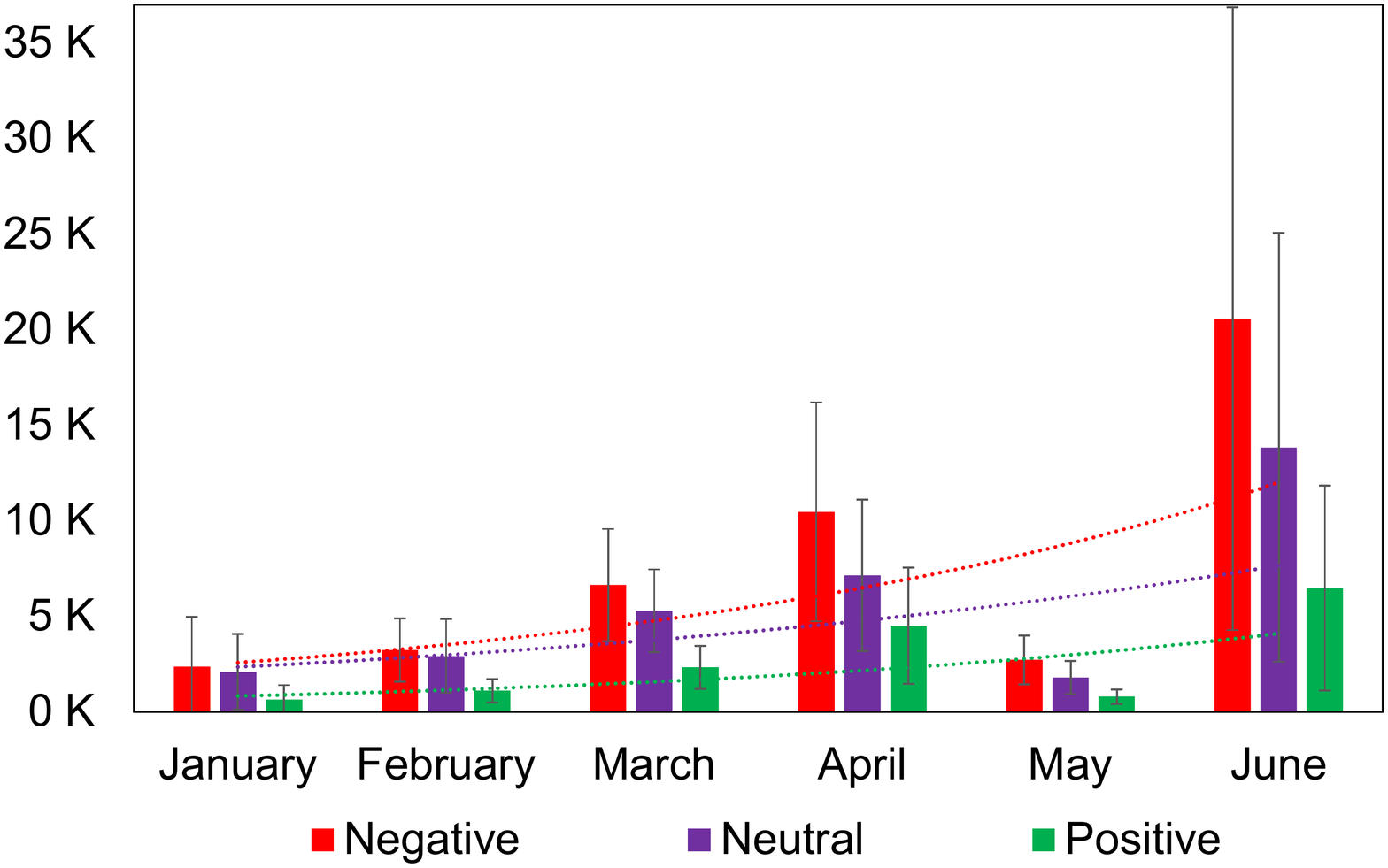}} \\ \multicolumn{2}{c}{(d) UK} \\
    {\includegraphics[width=0.5\linewidth]{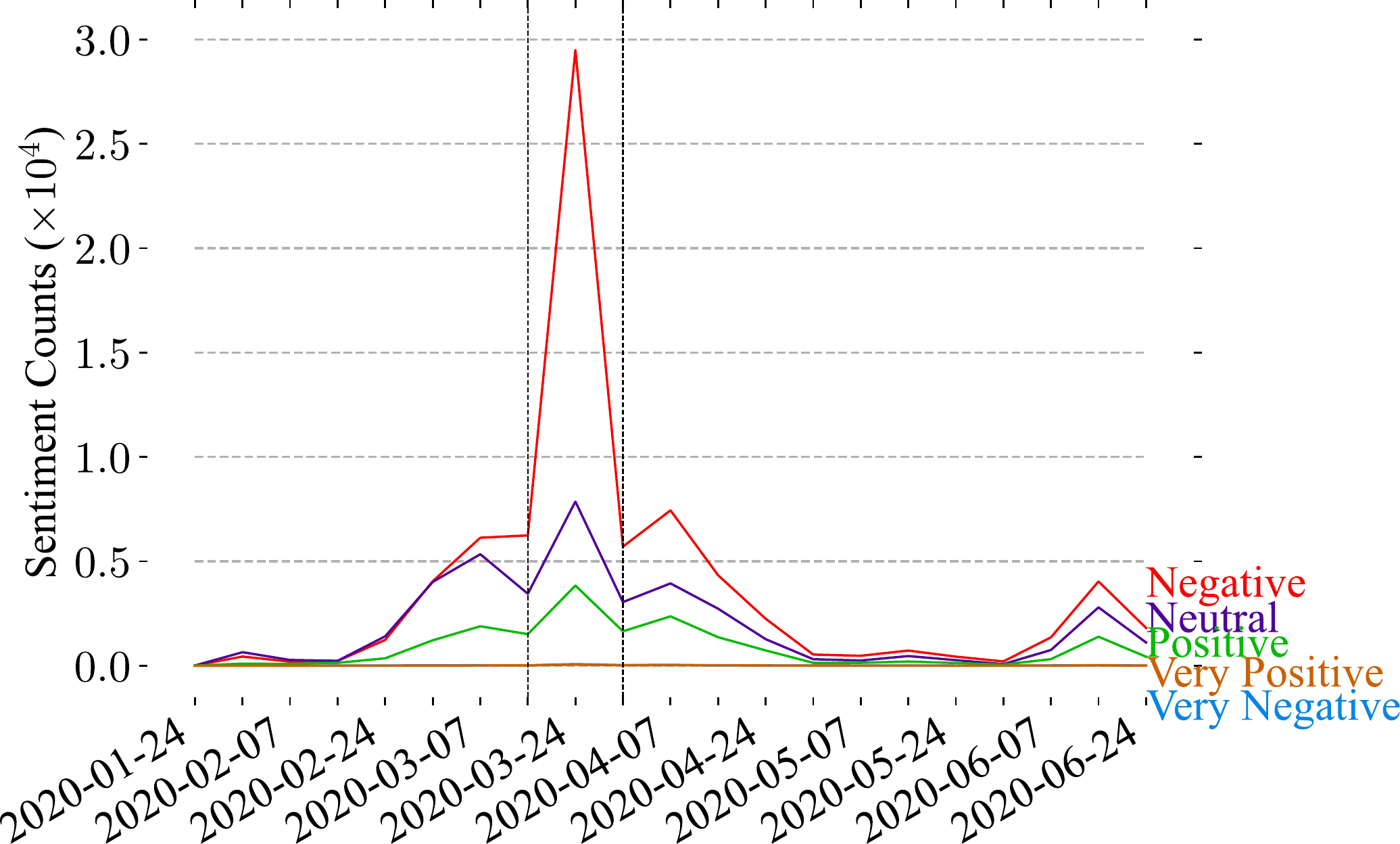}} & {\includegraphics[width=0.5\linewidth]{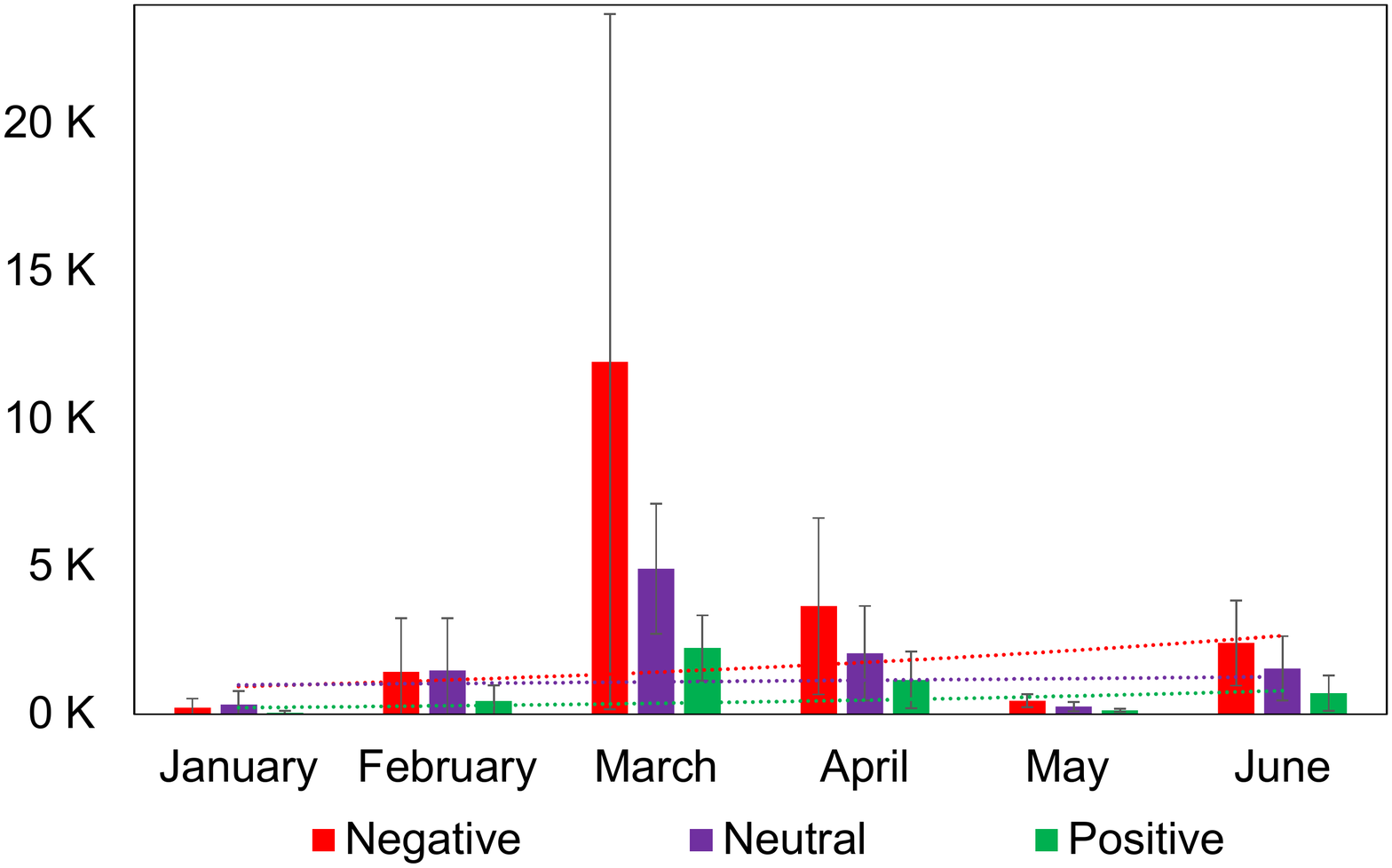}} \\ \multicolumn{2}{c}{(e) Italy} \\
\end{longtable} 
\captionsetup{justification=justified}
\captionof{figure}{Weekly and monthly sentiment analysis of most discussed countries on Twitter based on five sentiment classes (very negative, negative, neutral, positive, very positive). For weekly analysis (right side), spikes are highlighted with vertical black dashed lines, and trends are illustrated in the monthly analysis (left side) after merging the very positive and very negative to their corresponding classes.}
\label{fig:TwitterCountriesSentiment}
\addtocounter{table}{-1}%

\paragraph{Instagram} Figure \ref{fig:InstagramCountriesSentiment} illustrates the sentiment analysis of posts mentioning the top-5 countries, i.e., China, India, USA, Italy, and Spain, using our Instagram dataset. Figure \ref{fig:InstagramCountriesSentiment} shows that there are few posts mentioning the countries on Instagram. Based on the analysis, there are fluctuations of reactions as shown in Figure \ref{fig:InstagramCountriesSentiment}, therefore, there is no dominant sentiment type over time. Nevertheless, there are more occurrences of positive posts in the USA and Italy charts (see Figures \ref{fig:InstagramCountriesSentiment} (c) and (d)).

\begin{longtable}{cc}
    {\includegraphics[width=0.45\linewidth]{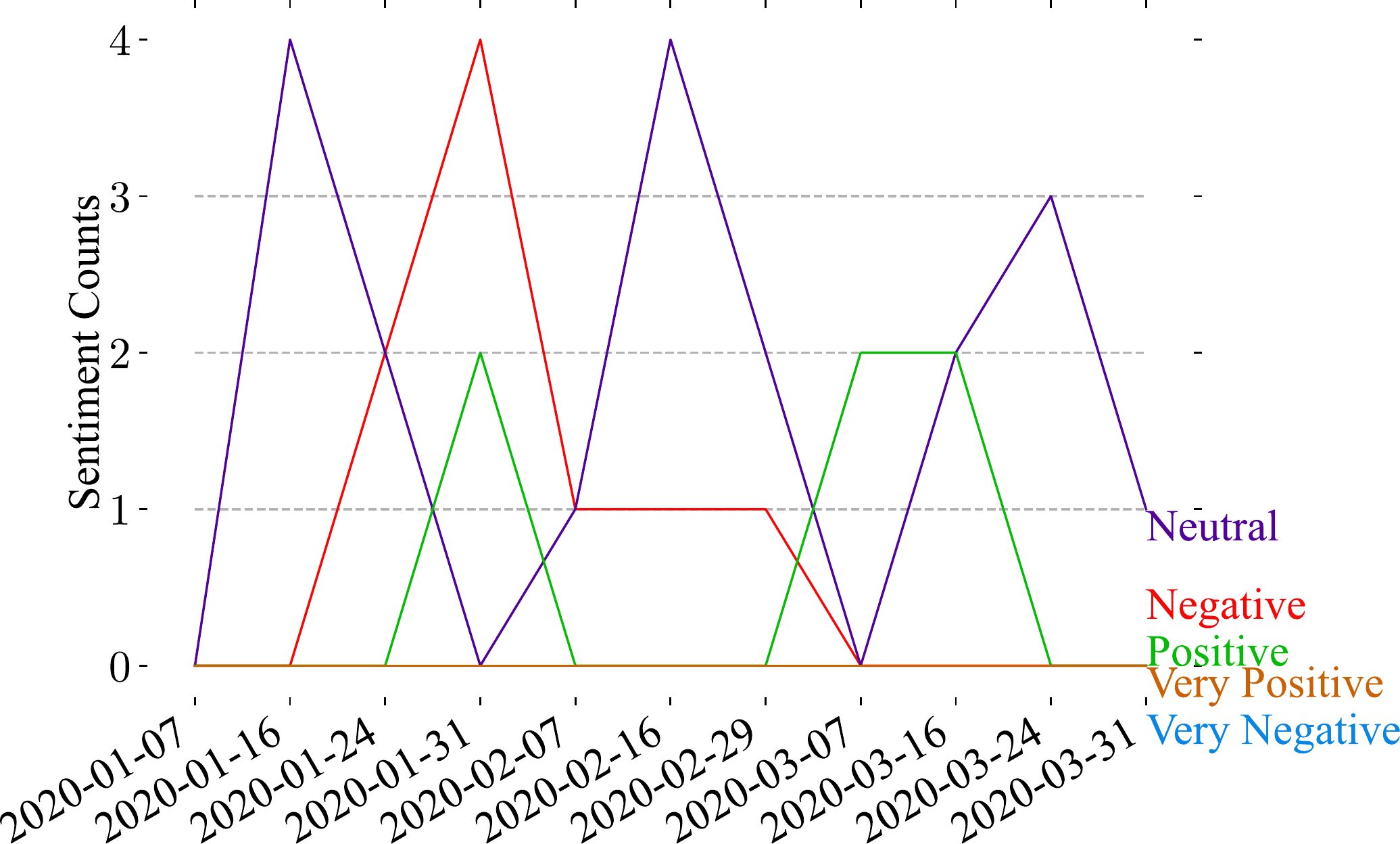}} & {\includegraphics[width=0.45\linewidth]{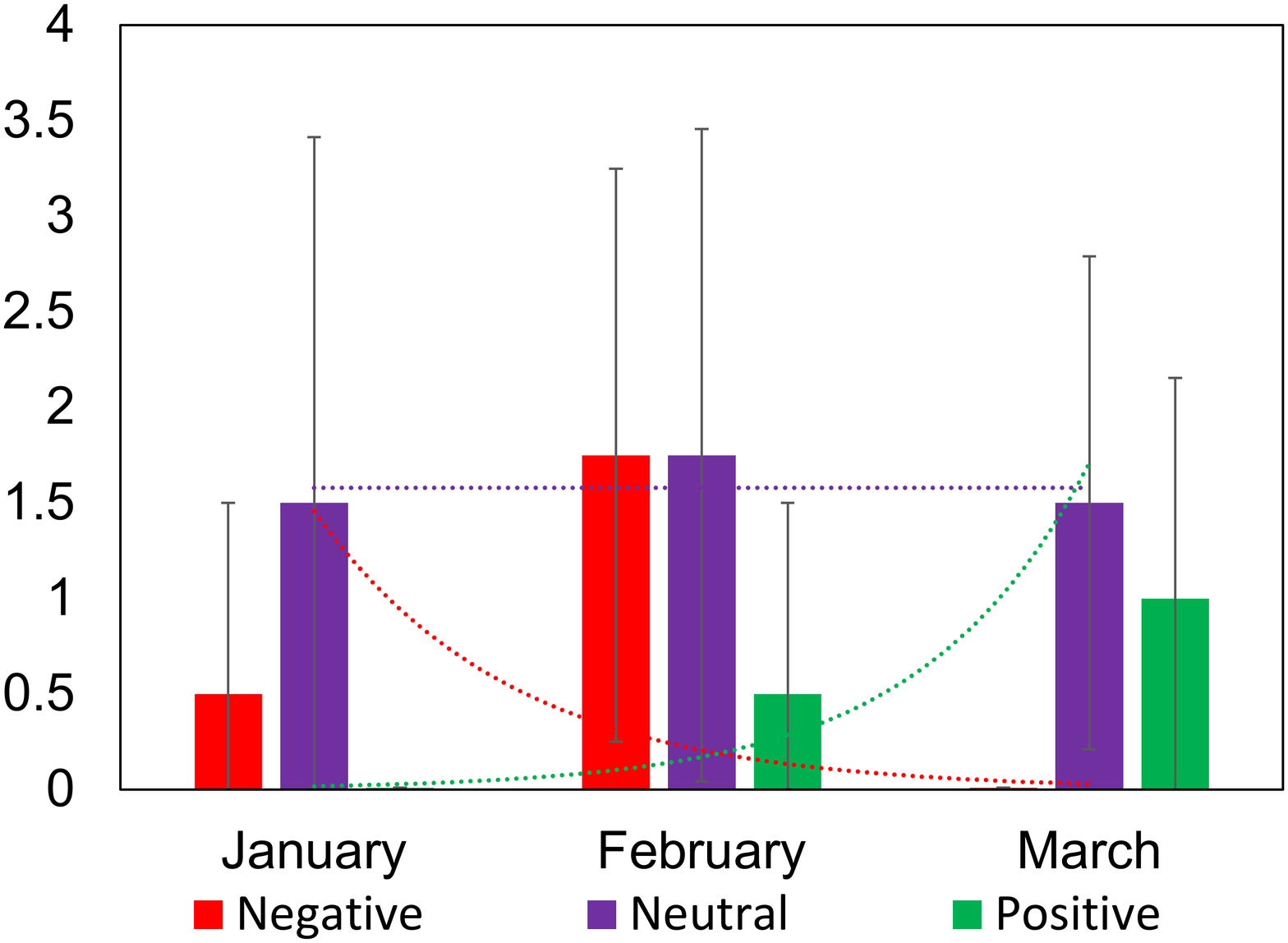}} \\ \multicolumn{2}{c}{(a) China} \\
    {\includegraphics[width=0.45\linewidth]{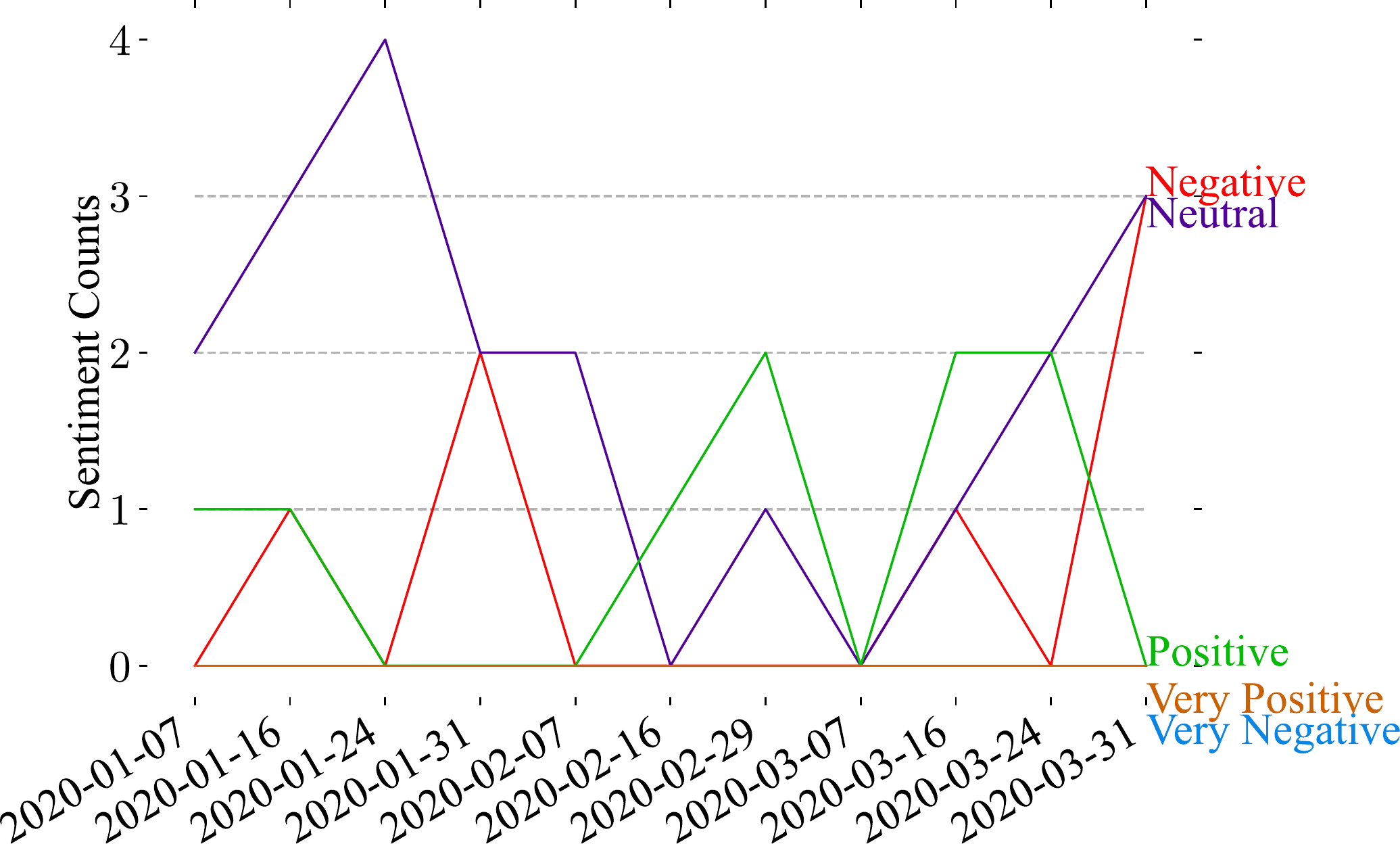}} & {\includegraphics[width=0.45\linewidth]{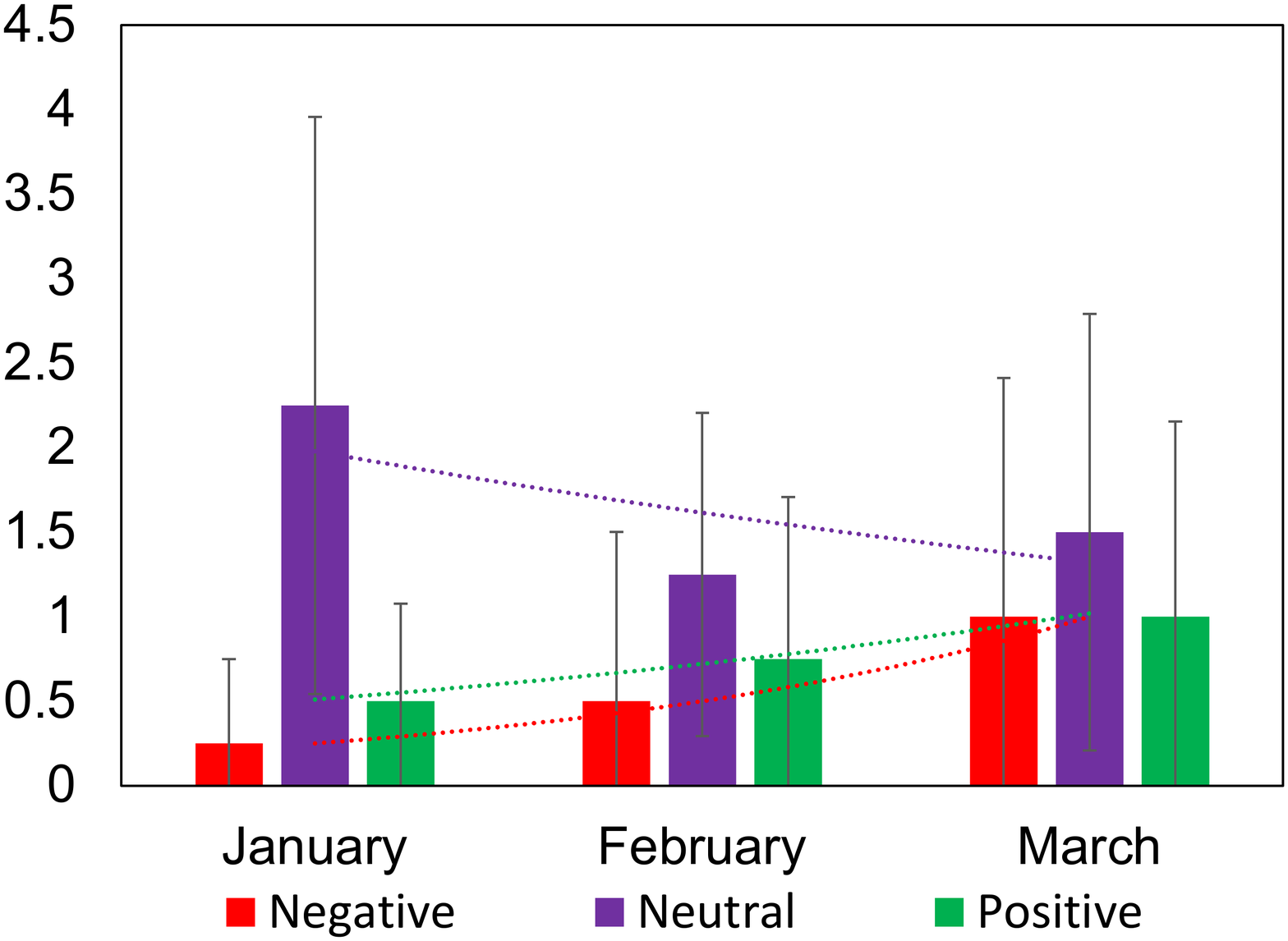}} \\ \multicolumn{2}{c}{(b) India} \\
    {\includegraphics[width=0.45\linewidth]{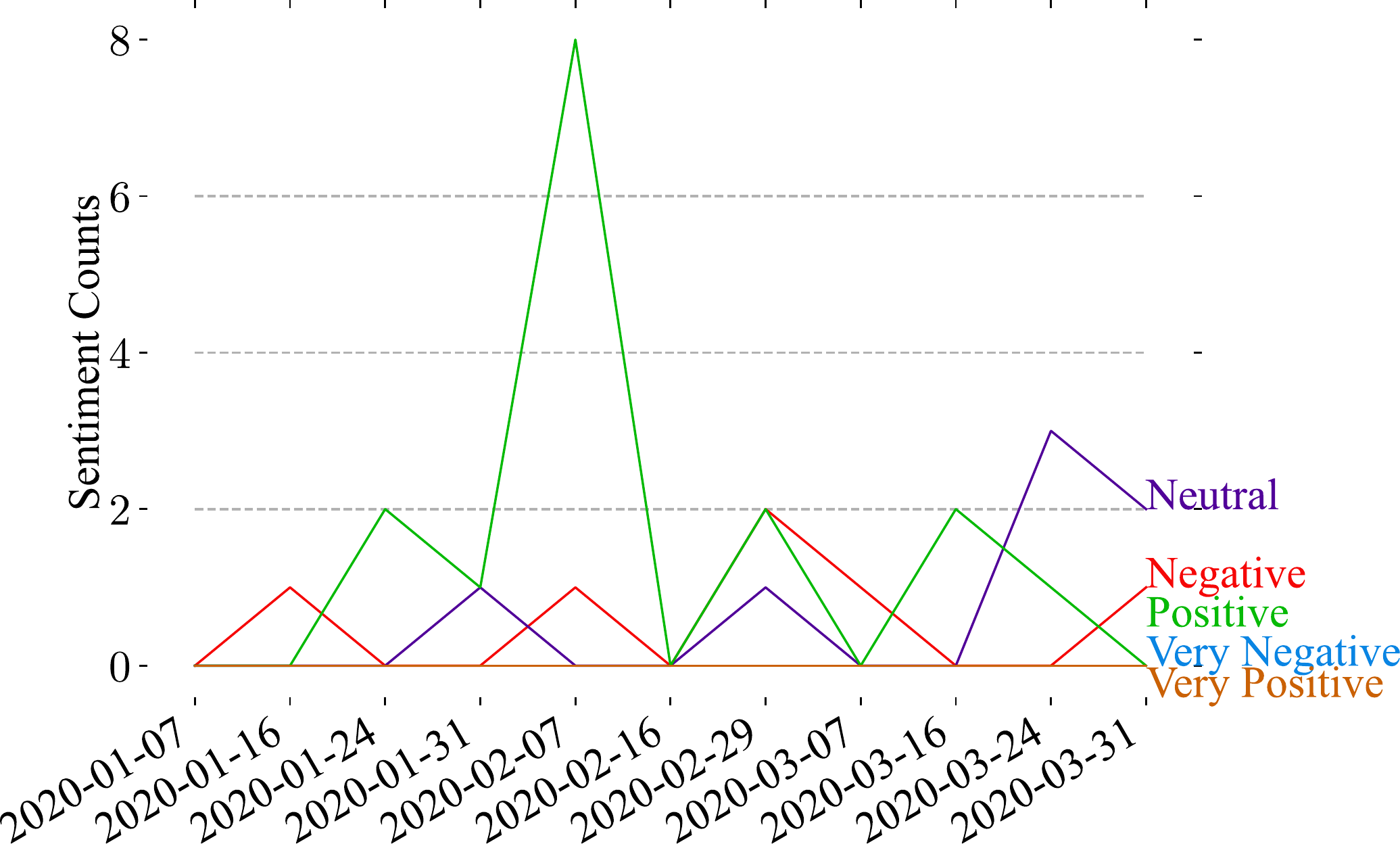}} & {\includegraphics[width=0.45\linewidth]{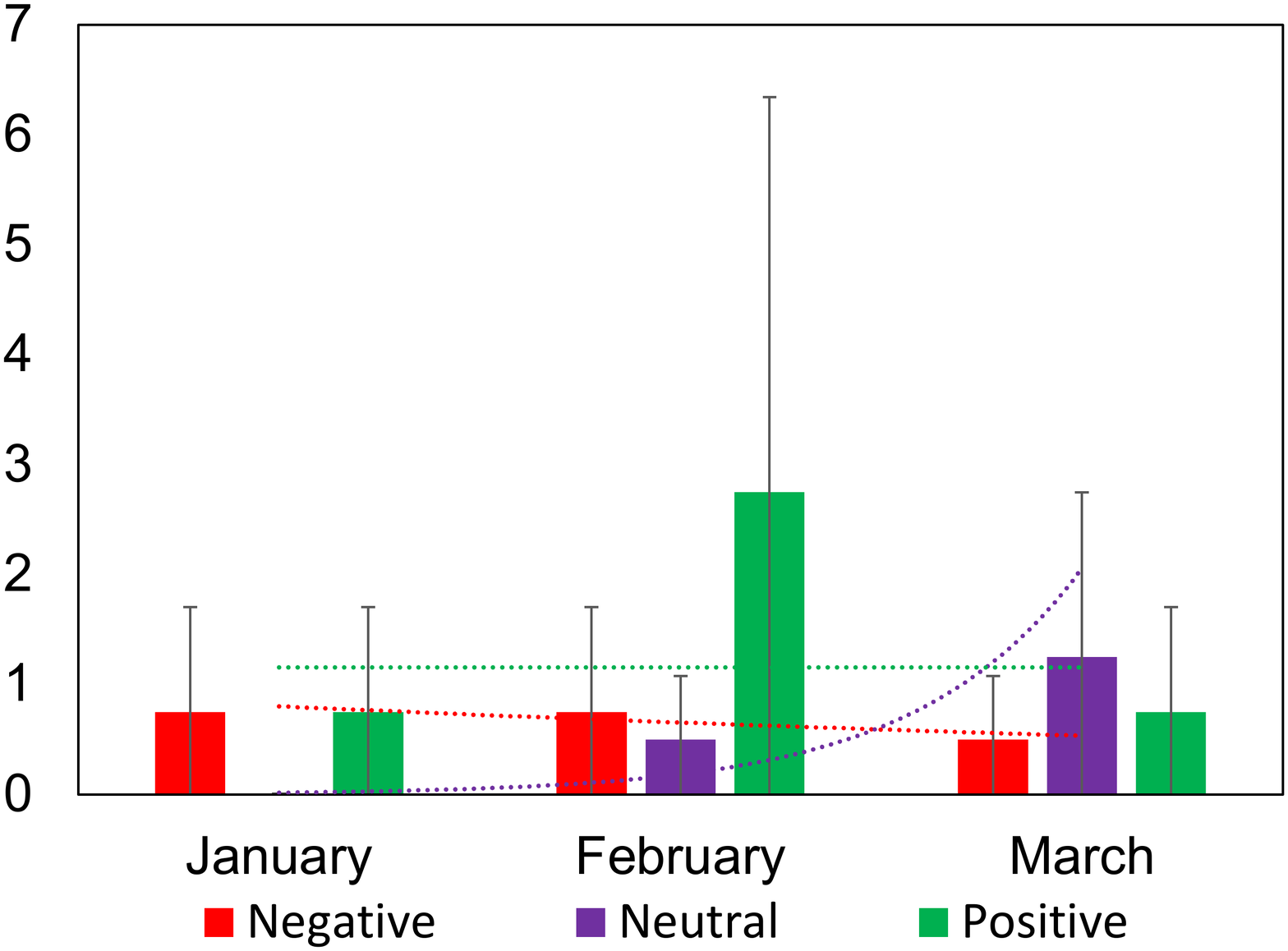}} \\ \multicolumn{2}{c}{(c) USA} \\
    {\includegraphics[width=0.45\linewidth]{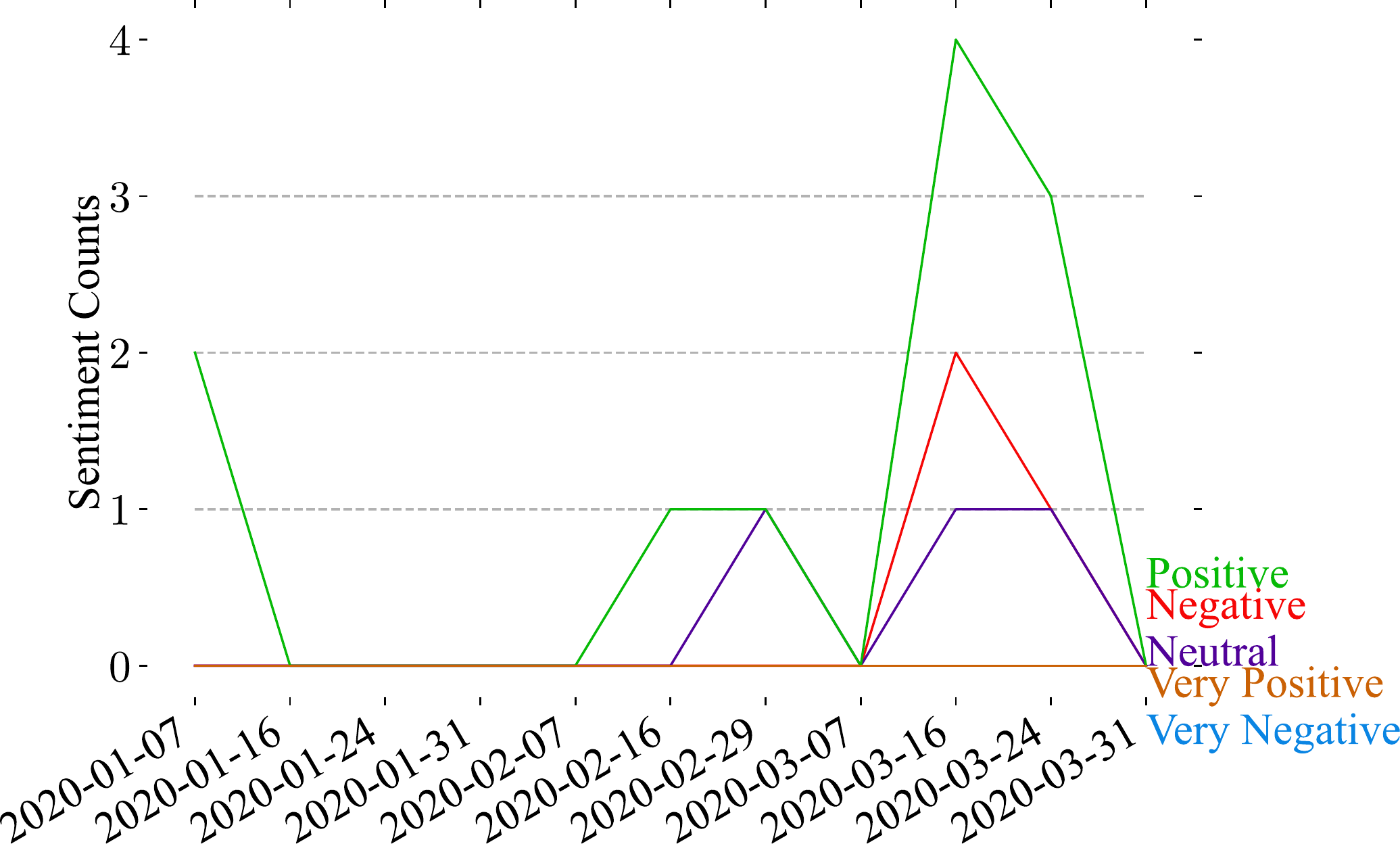}} & {\includegraphics[width=0.45\linewidth]{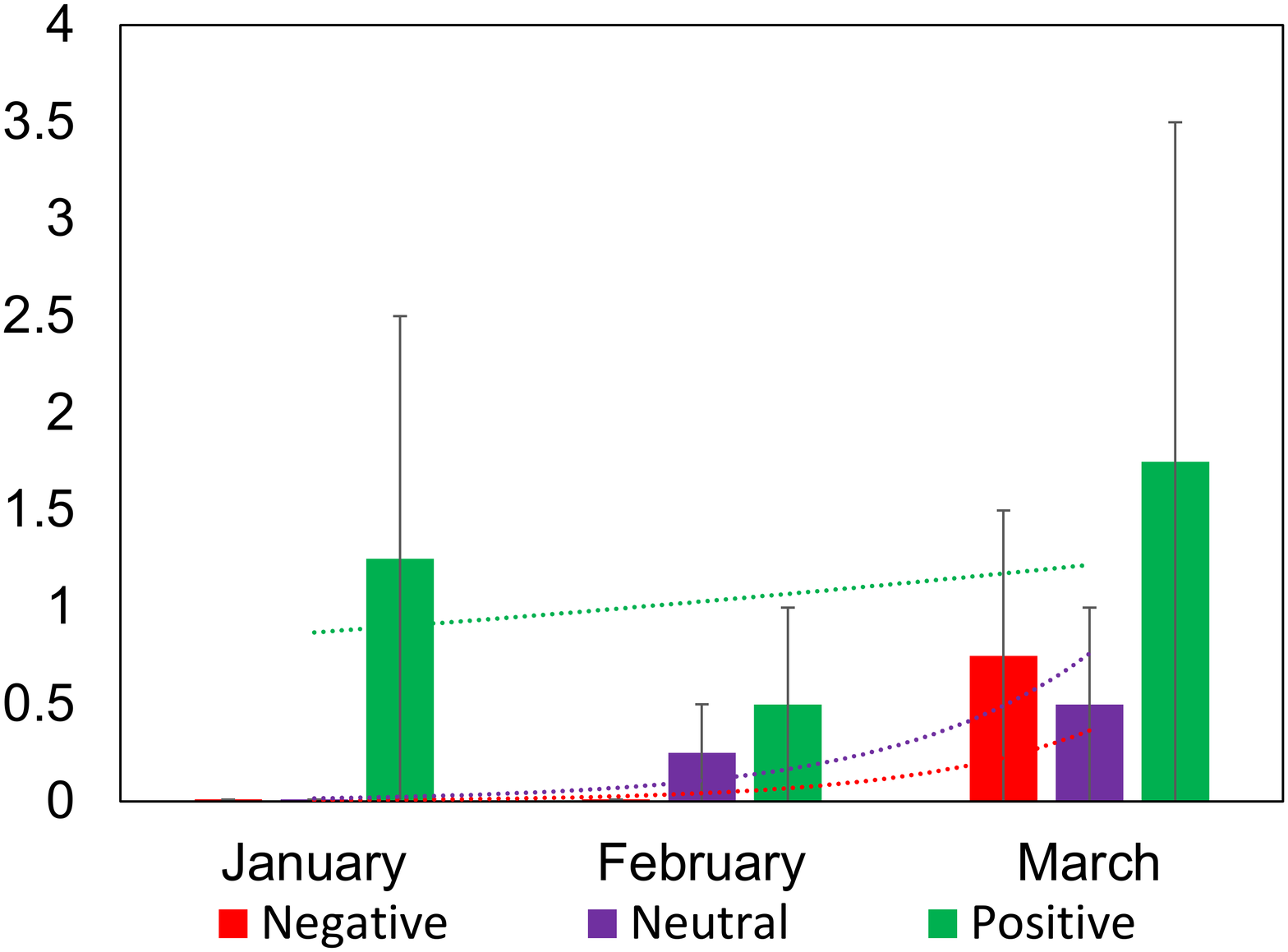}} \\ \multicolumn{2}{c}{(d) Italy} \\
    {\includegraphics[width=0.45\linewidth]{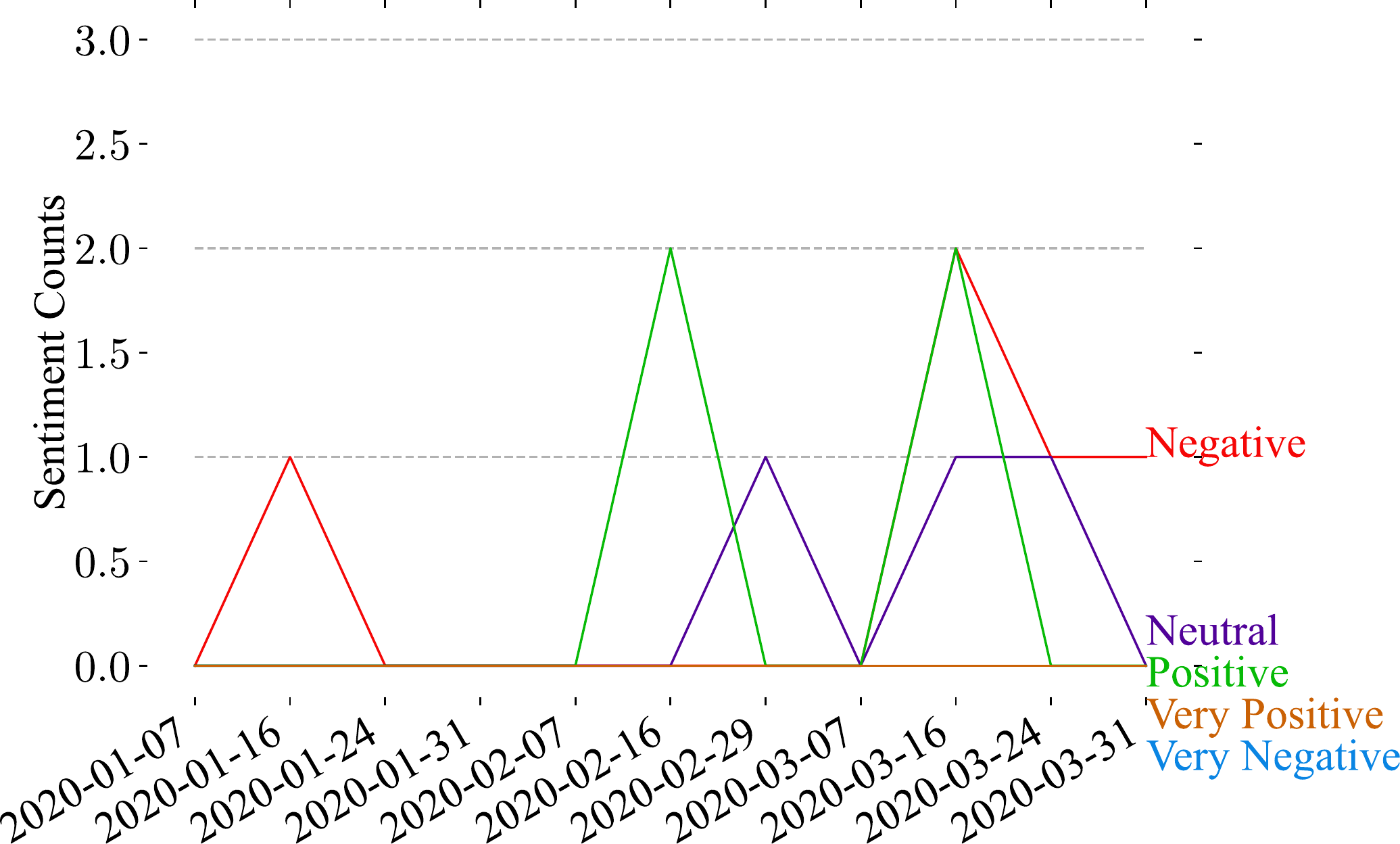}} & {\includegraphics[width=0.45\linewidth]{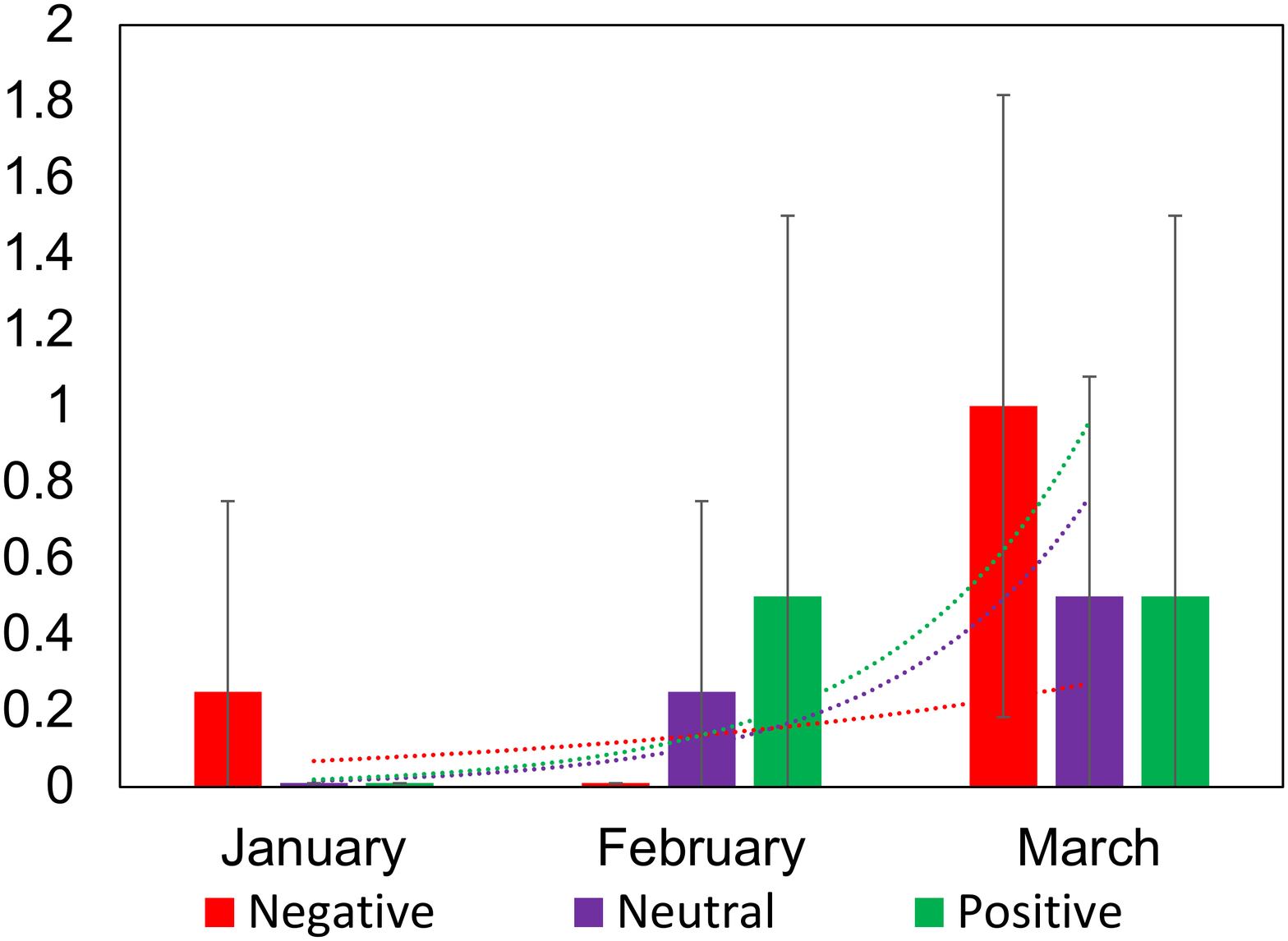}} \\ \multicolumn{2}{c}{(e) Spain} \\
\end{longtable} 
\captionsetup{justification=justified}
\captionof{figure}{Weekly and monthly sentiment analysis of most discussed countries on Instagram based on five sentiment classes (very negative, negative, neutral, positive, very positive). For weekly analysis (right side), spikes are highlighted with vertical black dashed lines, and trends are illustrated in the monthly analysis (left side) after merging the very positive and very negative to their corresponding classes.}
\label{fig:InstagramCountriesSentiment}
\addtocounter{table}{-1}%

\subsection{Topic Modeling on Top Countries} 
\setlength{\parindent}{3ex}
To further analyze the data of the top-5 countries, we conducted a topic modeling task to identify the major topics discussed in social media related to these countries. Identifying country-specific topics enables understanding how the people of these countries perceive the pandemic and to what extent their perception differs across localities. To this end, we employed LDA to extract country-specific topics in the top-5 mentioned countries in our dataset. Using the default implementation and settings of LDA in the Gensim LDA MultiCore model \cite{rehurek_lrec}, we explored various LDA models to extract a different number of topics ranging from 2 to 30. For our analysis, we selected the best-performing models, i.e., the models with the highest coherence score, to extract and identify the topics. The results of our study show the emergence of a variety of topics, such as safety messages, work, travel, COVID-19-related news, government response, emergency funding, and prevention measures, which are similar findings from Sharma et al. \cite{Sharma2020} and  Ordun et al. \cite{Ordun2020}.

\begin{figure}[h]
    \centering
    \captionsetup{justification=justified}
    \includegraphics[width=\textwidth]{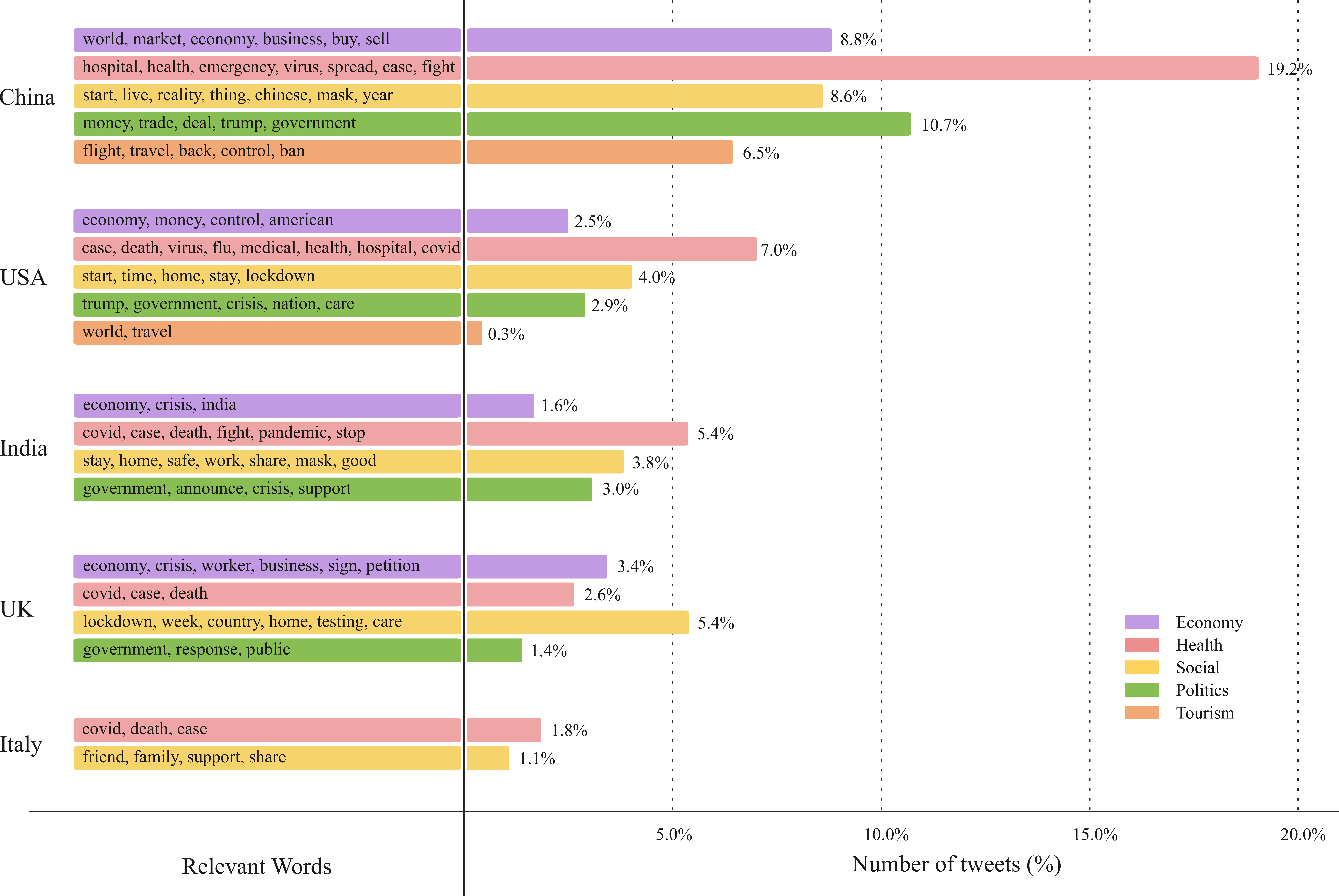}
    \caption{Topics generated for each country and the number of tweets belonged to each topic (Twitter)}
    \label{fig:TwitterCountryTopic}
\end{figure}

\paragraph{Twitter} Figure \ref{fig:TwitterCountryTopic} depicts the result of the topic modeling task on the Twitter dataset. The results show that health-related topics are the most-discussed topics across countries, except for the UK, where most tweets were related to the social category. The health-related tweets mentioning China constitute 19.2\% of the total tweets, indicating that the people’s interest in China’s health situation, including hospitals, virus spreads, number of cases, and how the Chinese handle the COVID-19 pandemic. Common concerns and trends in the health-related topics across all countries include COVID-19-related news, such as virus cases, the number of deaths, and virus spread. This high popularity of health-related topics indicates that people concern about the COVID-19 pandemic. In terms of the economy-related category, which constitutes 16.3\% of the total tweets in our dataset for the top-5 countries, the trend was on the virus’s economic aspects and implications. While our analysis highlighted more than four topics for tweets mentioning China, the USA, India, and the UK, only two topic categories were extracted from tweets mentioning Italy during the data collection period. Tweets mentioning Italy are related to health and social categories with 1.8\% and 1.1\% percentage of the total dataset, respectively. More specifically, tweets mentioning Italy focused on COVID-19 cases, deaths, and social aspects of life during the pandemic. 

Generally, the social-related tweets across different countries indicate the users’ attempt to inform others about prevention measures, such as staying at home, caring, supporting, sharing, wearing a mask, etc.

\begin{figure}[h] 
    \centering
    \captionsetup{justification=justified}
    \includegraphics[width=\textwidth]{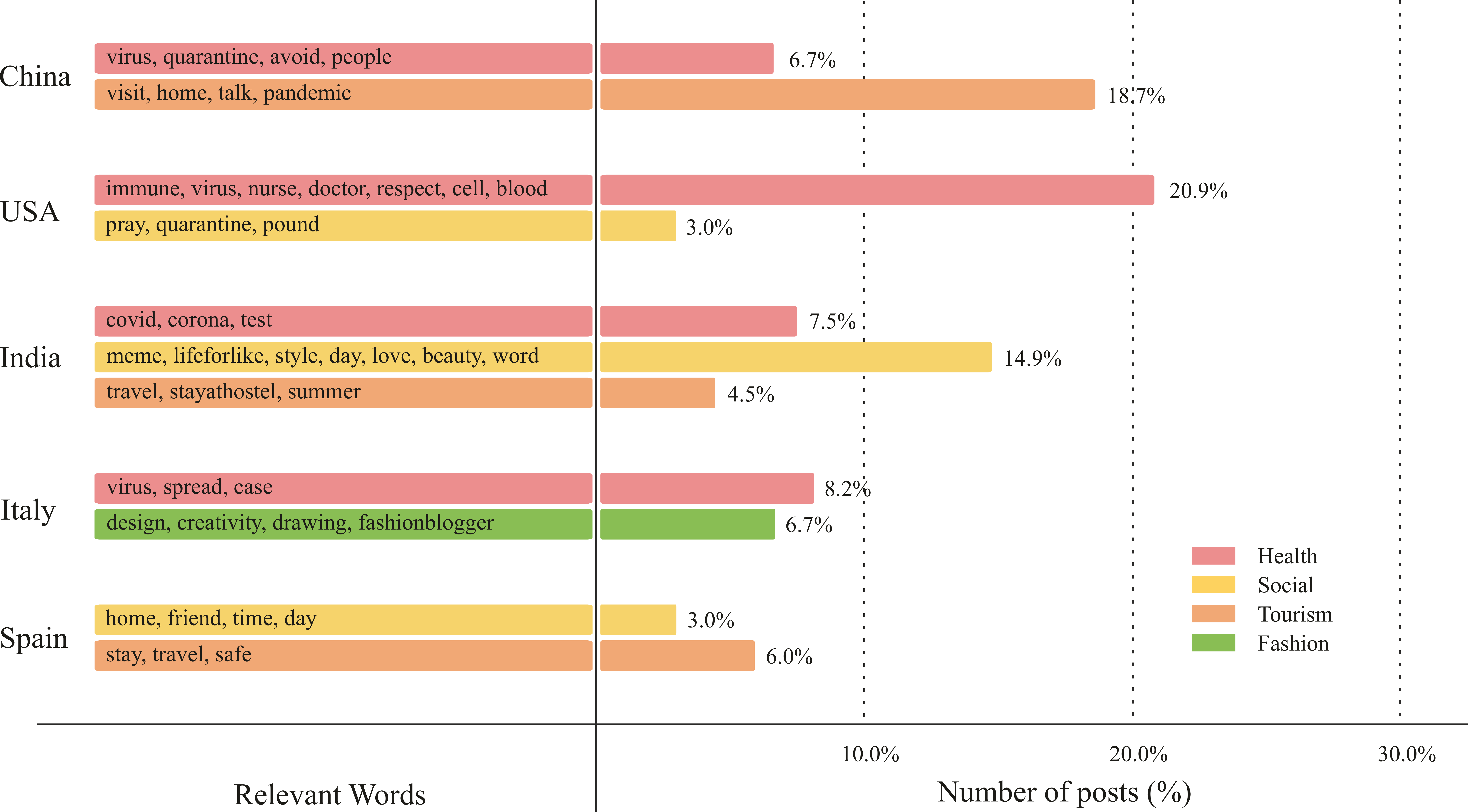}
    \caption{Topics generated for each country and the number of posts belonged to each topic (Instagram)}
    \label{fig:InstagramCountryTopic}
\end{figure}

\paragraph{Instagram} The result of the topic modeling analysis on the Instagram dataset is displayed in Figure \ref{fig:InstagramCountryTopic}.  The dominant topic categories across countries include health, social, fashion, and tourism, distributed with the percentages 43.3\%, 20.9\%, 6.7\%, and 29.2\%, respectively. Health-related posts are prominent with mentions of countries such as China, the USA, India, and Italy (as shown in Figure \ref{fig:InstagramCountryTopic}).  
Similar to trends from the Twitter dataset, users express their concern about the number of virus cases spread across different countries. Moreover, Figure \ref{fig:InstagramCountryTopic} shows that travel-related topics are the second most discussed topics in Instagram posts mentioning the top-5 countries with a total of 29.2\% of the entire dataset.

\subsection{Word2vec Analysis of Top Countries}

Our exploration of social media topics has shown that people expressed interest in aspects and issues in specific countries more than others. In previous sections, we discussed the topics mentioning the top-5 countries in social media. This section aims to identify the temporal patterns of used terms when mentioning the top-5 countries. It is interesting to explore the frequent words associated with a specific country and whether such words persisted or shifted during the pandemic. To this end, we apply a word embedding technique to investigate the context of terms associated with the top-5 countries in each month and for the entire collection period.

\begin{figure}
    \centering
    \captionsetup{justification=justified}
    \includegraphics[width=0.65\textwidth]{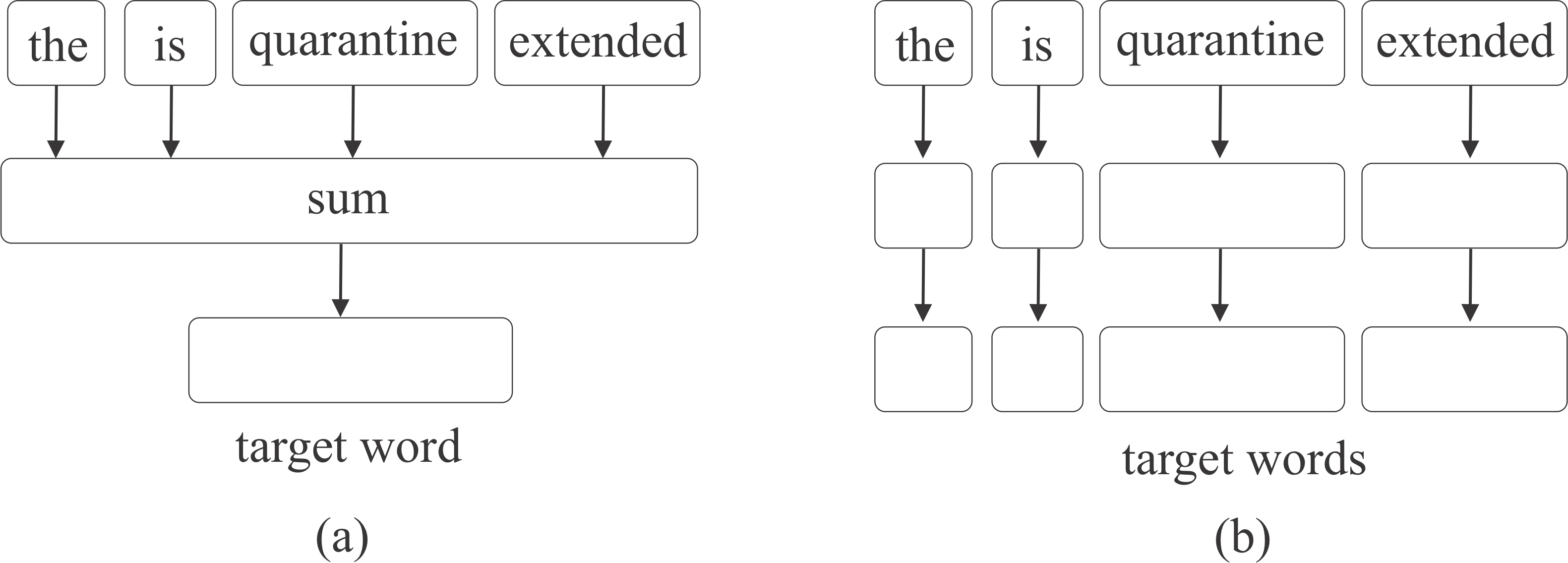}
    \caption{Word2Vec architecture for (a) bag of words (CBOW) and (b) Skip-gram algorithms}
    \label{fig:Word2Vec}
\end{figure}

\paragraph{Methods} We utilize the Word2Vec model that adopts a shallow neural network to embed words for the analysis. Word2vec has two main algorithms: a bag of words (CBOW) and a skip-gram (see Figure \ref{fig:Word2Vec}). The difference between those algorithms is that CBOW is used to predict a target word from a context, while skip-gram predicts a target context from a given word. For our implementation of the Word2Vec model, we used the Gensim library \cite{rehurek_lrec} with the default settings including hyperparameters such as \textit{window, size, sample, alpha, min-alpha, negative}, and \textit{workers}. However, we set the \textit{min\_count} hyperparameter to 20, which means that the words with occurrence less than 20 are ignored.

Word2Vec captures different degrees of similarity between words, and it uses vector arithmetic to reproduce syntactic and semantic patterns. Therefore, we use it to identify key terms and top countries on both social networks. 
Schild et al. \cite{Schild2020} used a similar technique, where the authors investigated the spread of Sinophobic slurs in social networks based on specific keywords such as “\textit{china}”, “\textit{chinese}”, “\textit{virus}” etc. The difference between the work of Schild et al. \cite{Schild2020}, and ours is that we use word embeddings to explore the dynamics and temporal patterns of used words in the social network over time, considering their association with different countries.

\begin{figure}
    \centering
    \captionsetup{justification=justified}
    \includegraphics[width=\textwidth]{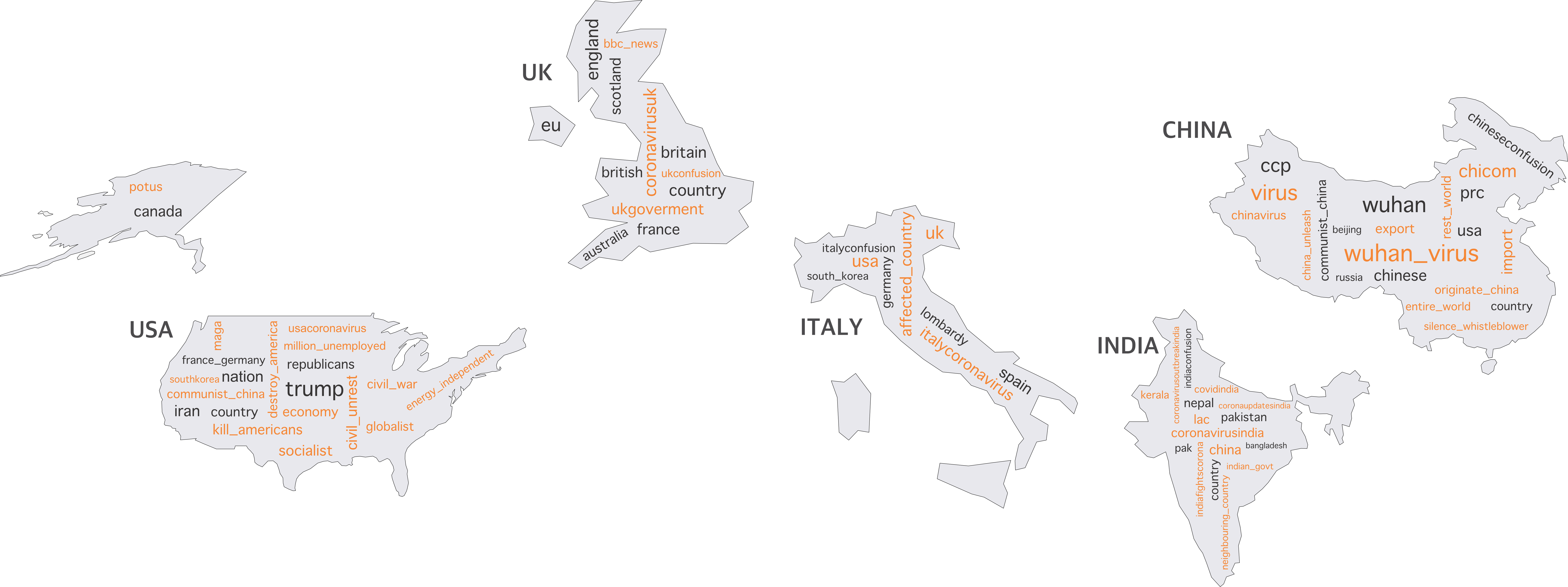}
    \caption{(Twitter) Words that are related to top countries. Black words represent words that remained unchanged between January and June. Orange words represent words that shifted towards the countries during the same period.}
    \label{fig:TwitterRelatedWords}
\end{figure}

\begin{figure}
    \centering
    \captionsetup{justification=justified}
    \includegraphics[width=\textwidth]{ 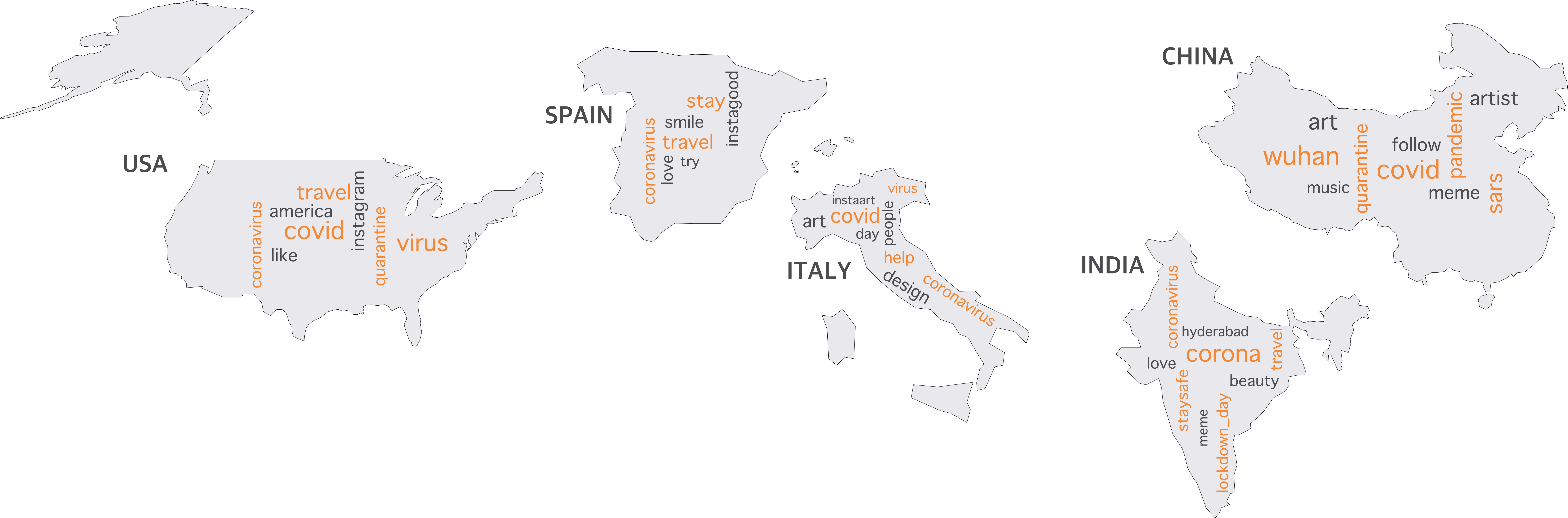}
    \caption{(Instagram) Words that are related to top countries. Black colored words represent words that remained unchanged between January and March. Orange colored words represent words that shifted towards the countries during the same period.}
    \label{fig:InstagramRelatedWords}
\end{figure}

The results of our analysis using Twitter and Instagram datasets are shown in Figures \ref{fig:TwitterRelatedWords} and \ref{fig:InstagramRelatedWords}, respectively. The figures show the words colored in black and orange. The black-colored words represent persistent words through the data collection period, and the orange-colored words represent trending words that shifted from January to June. Information about the most frequent words on both social networks is given in Tables \ref{tab:TwitterWords} and \ref{tab:InstagramWords}, respectively.

\begin{table}
\centering
\captionsetup{justification=justified}
\captionof{table}{Distribution of most frequent words of the Twitter dataset (\%)}
\label{tab:TwitterWords}
\resizebox{\linewidth}{!}{%
\begin{tabular}{l|cccccccccc}
\multicolumn{1}{c|}{ \textit{\textbf{Months}} } &  \textbf{\textit{covid} } &  \textbf{\textit{coronavirus} } &  \textbf{\textit{people} } &  \textbf{\textit{china} } &  \textbf{\textit{pandemic} } &  \textit{\textbf{virus}}  &  \textbf{\textit{lockdown} } &  \textit{\textbf{case}}  &  \textbf{\textit{corona} } &  \textit{\textbf{like}}  \\ 
\hline
 \textit{\textbf{January}}  & 0 & 8.20 & 4.12 & 20.00 & 0.63 & 9.06 & 0.42 & 8.87 & 1.12 & 3.54 \\
 \textbf{\textit{February} } & 1.60 & 14.30 & 9.90 & 42.89 & 2.35 & 17.52 & 0.59 & 10.80 & 9.22 & 11.52 \\
 \textit{\textbf{March}}  & 28.56 & 37.16 & 28.55 & 18.78 & 16.42 & 27.35 & 10.61 & 23.20 & 27.21 & 27.95 \\
 \textit{\textbf{April}}  & 38.21 & 25.89 & 20.74 & 8.31 & 23.45 & 14.94 & 21.00 & 22.48 & 12.33 & 19.10 \\
 \textit{\textbf{May }}  & 8.89 & 4.95 & 11.78 & 3.63 & 17.52 & 9.33 & 26.16 & 8.15 & 12.05 & 10.99 \\
 \textit{\textbf{June}}  & 22.72 & 9.50 & 24.90 & 39.63 & 39.63 & 21.80 & 41.23 & 26.50 & 38.07 & 26.89
\end{tabular}
}
\end{table}

\begin{table}
\centering
\captionsetup{justification=justified}
\captionof{table}{Distribution of most frequent words of the Instagram dataset (\%)}
\label{tab:InstagramWords}
\resizebox{\linewidth}{!}{%
\begin{tabular}{c|cccccccccc}
 \textit{\textbf{Months}}  &  \textbf{\textit{coronavirus} } &  \textbf{\textit{covid} } &  \textit{\textbf{love}}  &  \textit{\textbf{face}}  &  \textbf{\textit{follow} } &  \textbf{\textit{quarantine} } &  \textbf{\textit{stayhome} } &  \textit{\textbf{time}}  &  \textbf{\textit{corona} } &  \textit{\textbf{virus}}  \\ 
\hline
 \textit{\textbf{January}}  & 15.29 & 8.94 & 20.45 & 22.09 & 38.99 & 12.16 & 12.72 & 11.98 & 26.54 & 18.45 \\
\multicolumn{1}{l|}{ \textbf{\textit{February} }} & 35.95 & 23.10 & 38.66 & 30.67 & 28.52 & 24.71 & 22.37 & 17.96 & 26.15 & 30.36 \\
\multicolumn{1}{l|}{ \textit{\textbf{March}} } & 48.76 & 67.96 & 40.90 & 47.24 & 32.49 & 63.14 & 64.91 & 70.06 & 47.31 & 51.19
\end{tabular}
}
\end{table}

\paragraph{Twitter} Figure \ref{fig:TwitterRelatedWords} shows the color-coded words that persisted or shifted during the data collection period of Twitter posts mentioning different countries. We noticed a shift in using other words in association with different countries. For example, the term “\textit{wuhan\_virus}” was increasingly associated with mentioning “\textit{China}” over time, as well as other terms, such as “\textit{originate\_China}” and “\textit{chinavirus}”, indicating the people’s interest in communicating news about the virus as COVID-19 that was first documented in Wuhan, China. Other terms, such as “\textit{silence\_whistleblower}”, has shown a trend over time in tweets mentioning China. The phrase “\textit{silence\_whistleblower}” was in reference to the doctor who revealed the COVID-19 in China, e.g., one tweet says: “\textit{Sad: Chinese doctor who worked with late whistleblower dead from \#coronavirus}”. Our analysis also highlights terms that were associated with the USA, including terms that are not related to coronavirus but to events that occurred during the pandemic. For example, the terms “\textit{civil\_unrest}” and “\textit{civil\_war}” were associated with the death of George Floyd on May 25, 2020, and its consequences nationwide. Another trend was captured for the phrase “\textit{million\_unemployed}” which was more likely related to an article published in The New York Times \cite{NewYorkTimes} pointing out the unemployment numbers in the USA, specifically, mentioning that 51 million Americans were unemployed since the start of the COVID-19 pandemic. The following tweet is an example for this case: “\textit{Coronavirus: 3.3 million more unemployed in a week in the United States, a historic record …}”. Similar coronavirus-related trends were also obvious in association with other countries, such as India, the UK, and Italy.

\paragraph{Instagram} The result of word representations and trends using the Instagram posts is shown in Figure \ref{fig:InstagramRelatedWords}. The results show that some terms such as, “\textit{smile}”, “\textit{love}”, “\textit{instagood}”, “\textit{art}”, “\textit{follow}”, etc. remain unchanged within the data collection period. However, some other terms were shifted, such as “\textit{travel}”, “\textit{help}”, “\textit{stay\_safe}” across different countries. We also observed the trends for using terms related to coronavirus over time. Accordingly, this means that people started publishing posts related to COVID-19 indicating their attention to the pandemic’s spreading. Some of such posts show that Instagram users were trying to compare past diseases, e.g., severe acute respiratory syndrome (SARS), with COVID-19. This can be the reason for observing the keyword “\textit{sars}” in association with China, especially referencing the beginning of the SARS outbreak in China (November 2002, Xu et al. \cite{Xu2004}). 

\subsection{Locality Analysis of Posts Associated with Top Countries}

To study users' localities interacting on social media through posts mentioning the top-mentioned countries, we further investigated the posts and locations of users showing interest in different topics associated with these countries. Indeed, users worldwide can share an interest in a variety of topics across other countries. Therefore posts on social media mentioning a specific country do not necessarily need to be originated from that country. In this section, we aim to identify the localities of posts mentioning the top countries.

\begin{figure}
    \centering
    \captionsetup{justification=justified}
    \includegraphics[width=0.8\textwidth]{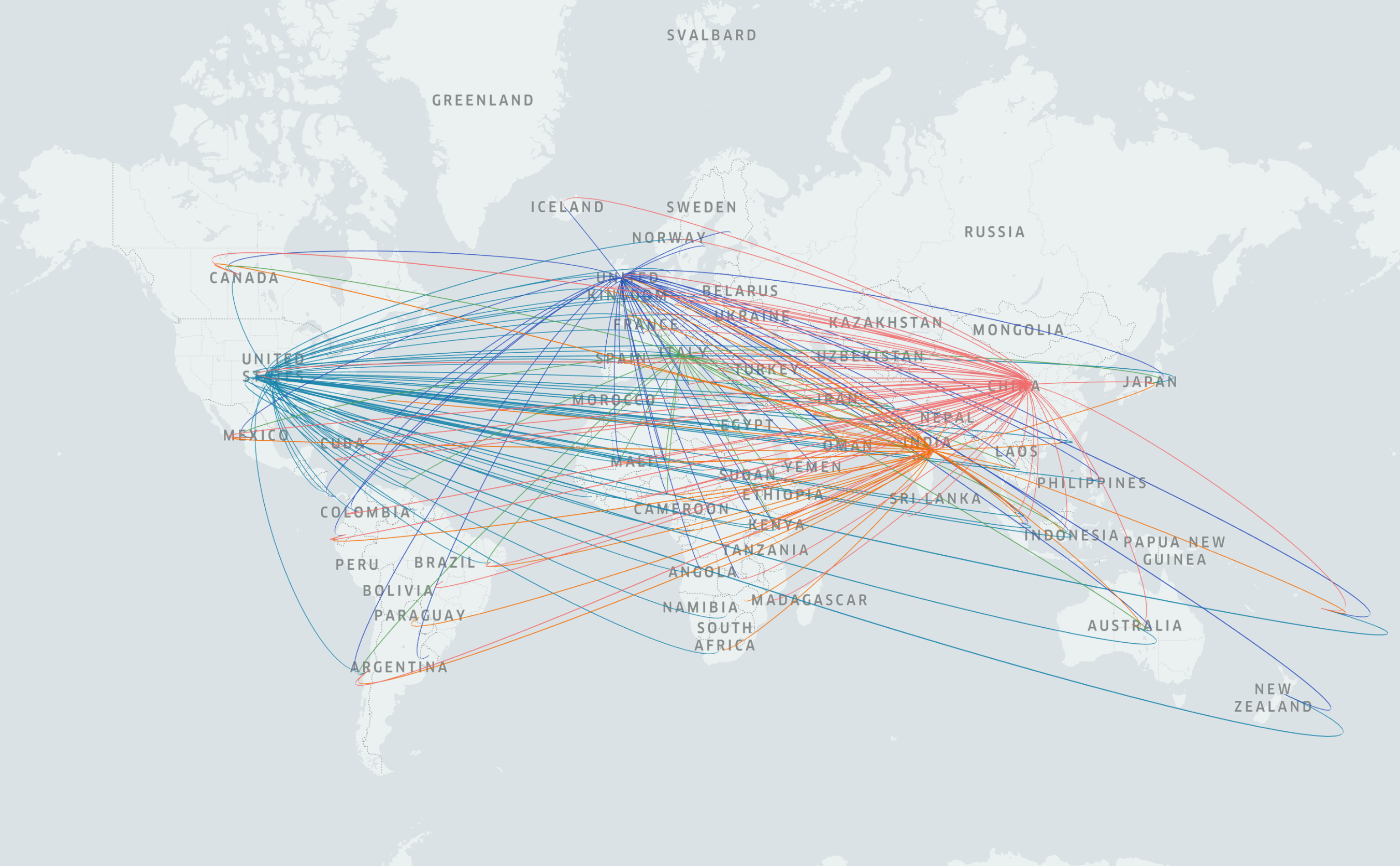}
    \caption{Countries where users published posts about China, India, the USA, Italy, and the UK on Twitter}
    \label{fig:TwitterMap}
\end{figure}

\paragraph{Twitter} Figure \ref{fig:TwitterMap} displays locations on a map pointing to the top countries, where the to connect lines mean that users from those locations published tweets about the top countries. China-related tweets include 66 countries, while tweets about the USA, UK, India, Italy are published from 64, 50, 44, and 26 countries, respectively.

Taking those number of countries into account (the numbers do not include the count of top countries), it can be said that China was mentioned by users from more places than other countries, while the USA is the second highest one. Since our results and analysis showed that the most discussed topics are related to the health category, we can deduce that most countries discussed the health situations in China. Even though a smaller number of countries mentions Italy, people had discussions only about its health and social causes. It should be underlined that people from 50 different locations focused on the UK's social parts while the health category is dominant for other top countries.

\begin{figure}
    \centering
    \captionsetup{justification=justified}
    \includegraphics[width=0.8\textwidth]{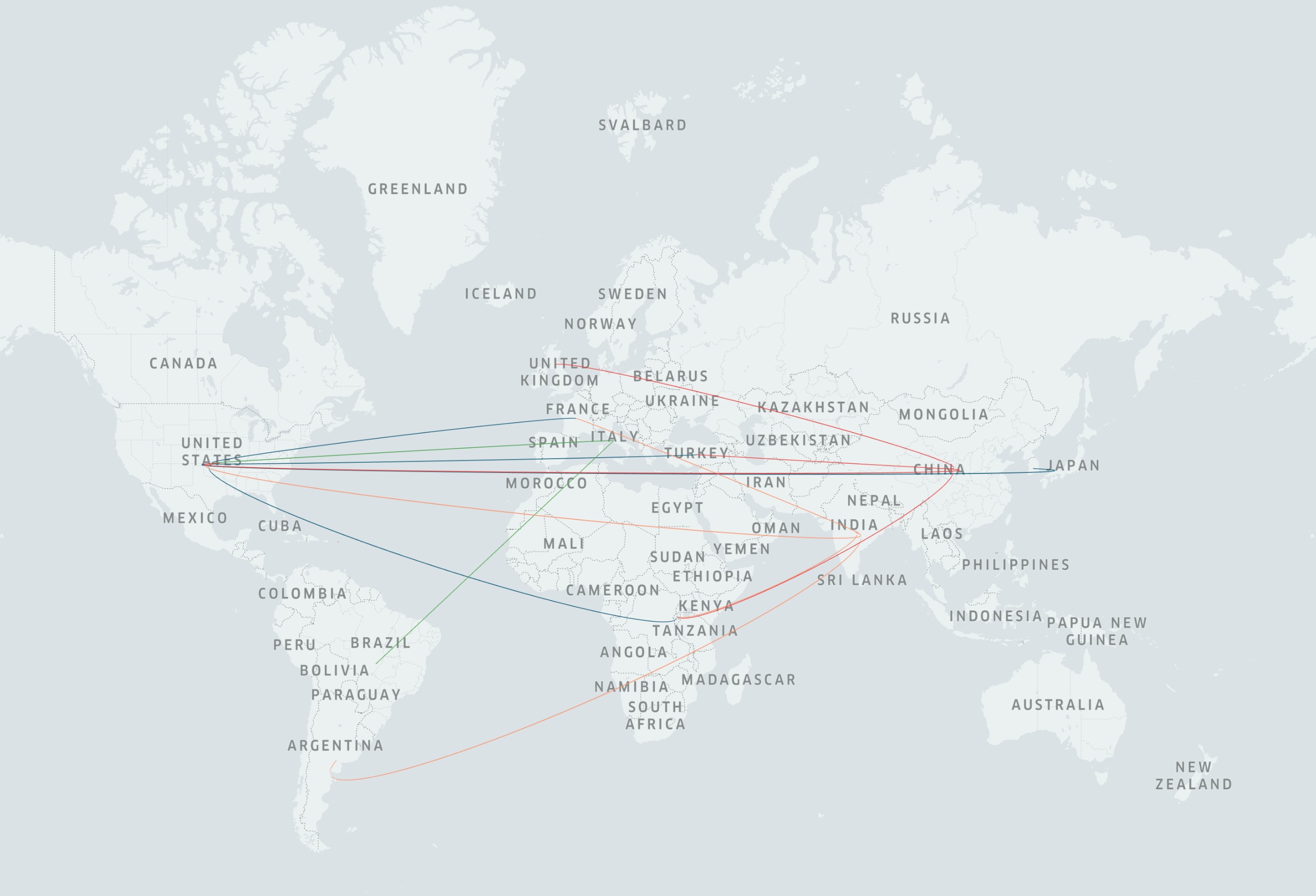}
    \caption{Countries where users published posts about China, India, the USA, Italy, and Spain on Instagram}
    \label{fig:InstagramMap}
\end{figure}

\paragraph{Instagram} Figure \ref{fig:InstagramMap} depicts the results of our locality analysis of posts associated with the top countries. The figure shows that there are less localities, where people mentioned the top countries in comparison to Twitter data. For example, Spain is not mentioned by any user from other countries, while China, the USA, and India have only four separate external locations where people published posts about them. Italy is discussed by only two other countries, i.e., Brazil and the USA. Users from Turkey, the UK, Rwanda, the USA, and China paid attention to topics that are related to travel. Health-related topics from the USA were the main point of interest to users from South Korea, France, Turkey, Rwanda, and the USA.

\section{Conclusion}

First, we identified people’s overall reactions that could be deduced from sentiment analysis. We found that users of both social networks published more neutral tweets and posts, but the number of negative tweets is also significantly high. Second, we implemented topic modeling on the datasets with distinct sentiment types (negative, neutral, positive) on specific periods that led to spikes. Third, we performed the country analysis to observe countries in discussion and found that China, the USA, India, the UK, and Italy are the countries that attracted more Twitter users’ attention. Interestingly, those countries, except for the UK, are the most discussed countries together with Spain by Instagram users. Fourth, by displaying each country’s sentiment analysis on both social media, we detected spikes on specific time periods and identified what topics led to those spikes. It can be inferred that topics related to economy, politics, health, social, and tourism are the reasons for those spikes on Twitter. In contrast, health, fashion, social, and tourism categories are superior on Instagram. We can conclude that on both social media platforms, health topics are governing over other categories based on the resultant Figures \ref{fig:TwitterCountryTopic} and \ref{fig:InstagramCountryTopic}. Fifth, we detected the words that remained or shifted towards top countries by carrying out the word2vec analysis. The results showed that COVID-19 affected top countries, and those countries experienced a shift of meaning. Thus, terms related to coronavirus are linked with the country names of Twitter and Instagram. Sixth, we measured the impact areas of top countries by identifying the location of publishing retweets and posts. According to our analysis results, users of 66 countries (Twitter) along with another four countries (Instagram) focused on China more than any other country. In conclusion, as our analysis provides a detailed insight for COVID-19 datasets of both social networks, this paper can be used to comprehend people’s reactions and their attention.

\section*{Acknowledgements}
This work was supported by the National Research Foundation of Korea(NRF) grant funded by the Korea government(MSIT) (No. 2021R1A2C1011198). 

\bibliographystyle{unsrtnat}
\bibliography{mybibfile}  






\end{document}